\documentclass[a4paper,fleqn,usenatbib]{mnras}

\usepackage{newtxtext,newtxmath}
\usepackage[T1]{fontenc}
\usepackage{ae,aecompl}
\usepackage{graphicx}	% Including figure files
\usepackage{amsmath}	% Advanced maths commands
\usepackage{enumitem}
\usepackage{rotating}
\usepackage{pdflscape}

\newcommand{\xmm}{\emph{XMM-Newton}}

\newcommand{\kms}{km\,s$^{-1}$}

\title[X-rays from WR21 and WR31]{Colliding winds in WR21 and WR31 - I. The X-ray view\thanks{Based on spectra obtained with \xmm }}

\author[Y. Naz\'e et al.]{Ya\"el~Naz\'e$^1$\thanks{F.R.S.-FNRS Senior Research Associate, email: ynaze@uliege.be}, Gregor Rauw$^1$, Rachel Johnson$^2$, Eric Gosset$^1$\thanks{F.R.S.-FNRS Research Director}, and Jennifer L. Hoffman$^2$ 
\\
$^1$ Groupe d'Astrophysique des Hautes Energies, STAR, Universit\'e de Li\`ege, Quartier Agora (B5c, Institut d'Astrophysique et de G\'eophysique), \\
All\'ee du 6 Ao\^ut 19c, B-4000 Sart Tilman, Li\`ege, Belgium\\
$^2$ Department of Physics and Astronomy, University of Denver, Denver, CO 80210 USA
}
\pubyear{2023}

\begin{document}
\label{firstpage}
\pagerange{\pageref{firstpage}--\pageref{lastpage}}
\maketitle

\begin{abstract}
WR21 and WR31 are two WR+O binaries with short periods, quite similar to the case of V444\,Cyg. The \xmm\ observatory has monitored these two objects and clearly revealed phase-locked variations as expected from colliding winds. The changes are maximum in the soft band (0.5--2.\,keV, variations by a factor 3--4) where they are intrinsically linked to absorption effects. The increase in absorption due to the dense WR wind is confirmed by the spectral analysis. The flux maximum is however not detected exactly at conjunction with the O star in front but slightly afterwards, suggesting Coriolis deflection of the collision zone as in V444\,Cyg. In the hard band (2.--10.\,keV), the variations (by a factor of 1.5--2.0) are much more limited. Because of the lower orbital inclinations, eclipses as observed for V444\,Cyg are not detected in these systems.
\end{abstract}

\begin{keywords}
stars: early-type -- stars: massive -- binaries: general -- X-rays: stars
\end{keywords}

\section{Introduction}
The category of Wolf-Rayet (WR) stars was defined on the basis of their peculiar spectra, composed of broad and strong emission lines. These lines are mostly associated to helium, nitrogen, carbon, and oxygen ions. Their relative strengths are used to classify these stars into the different WR subcategories (WN, WC, WO). These emissions do not arise at the photosphere of the stars, but in their very dense winds. Such winds can only be achieved for extremely massive stars still in the Hydrogen core-burning phase (the so-called WNLh stars, e.g. WR22, WR25) or in the last stages of evolution after envelope shedding (e.g. WR140).

When a WR star is paired with another massive star, their winds can collide, leaving a specific signature throughout the whole electromagnetic spectrum. In particular, X-ray emissions can be expected if the wind-wind collision creates a hot plasma ($>10^6$\,K, \citealt{che76,ste92}). The detection of bright X-rays from WR binaries, decades ago, thanks to $Einstein$ observations demonstrated that this indeed occurs \citep{pol87}. It must be noted, however, that not all WR binaries appear bright in X-rays  \citep{nazwr}. 

One specific characteristic of these colliding wind X-rays is their periodic variability. In eccentric binaries, the orbital separation $D$ between the stars changes, leading to changes in the intrinsic emission of the wind collision region. For purely adiabatic cases, a $1/D$ variation of that emission is expected \citep{ste92} and has indeed been observed, e.g. in WR25 \citep{gos07,pan14}. A different behaviour occurs in purely radiative collisions, see e.g. the linear variation with $D$ detected in HD\,5980 \citep{naz18}. Note that, in some cases, the collision may change nature with phase, and the resulting behaviour then appears more complex.

The emitted X-rays can be absorbed by the circumstellar material, which also leads to phase-locked changes. Indeed, the line-of-sight towards the collision zone may cross either wind and, if a strong contrast exists between them, large changes in the observed (soft, i.e. below 2\,keV) X-ray flux will be detected (e.g. a factor of four for $\gamma^2$\,Vel in $ROSAT$ data, \citealt{wil95}). In addition, absorption may also increase at phases near periastron in eccentric systems as the collision zone dives deep into the WR wind (e.g. increase by a factor of $\sim$4 in WR22, \citealt{gos09}). Again, both of these effects can combine, leading to more complex variations (e.g Melnick\,33Na, \citealt{bes22}). Finally, additional effects may also take place. For example, a disruption of the collision zone may occur at or near periastron, resulting in a lower X-ray emission at that phase, followed by a slow recovery as the orbital separation increases (e.g. $\eta$\,Car - \citealt{pan18}, WR21a - \citealt{gos16}, Melnick\,34 - \citealt{pol18}).

Establishing the X-ray properties is very useful as they are directly linked to the collision physics, the stellar properties (mass-loss rate, wind velocity...), and the binary geometry (separation, inclination...). In this context, as many phenomena may be at play, it is interesting to analyze specific systems which are subject to a limited number of effects. For example, \citet{lom15} analyzed the X-ray and spectropolarimetric light curves of V444\,Cyg (WN5+O6). This short-period WR binary ($P=4.21$\,d) possesses a circular orbit, eliminating all effects due to eccentricity. Its dedicated X-ray monitoring revealed changes in absorption, with a low value associated to phases at which the line-of-sight crosses the O-star wind (and a high one when the WR wind was involved). In addition, eclipses of the collision zone, due to the high inclination (78$^{\circ}$) of the system, were detected for the first time as well as a light curve asymmetry due to Coriolis distortion of the collision zone.

Using \xmm, we have now gathered data for two other short-period WR binaries with circular orbits: WR21 (WN5+O7V, $P=8.25$\,d) and WR31 (WN4+O8V, $P=4.83$\,d) - see \citet{fah}. The former system is formed of stars quite similar to those in V444\,Cyg, but with a longer period (hence a wider separation). The latter system has a period similar to that of V444\,Cyg, but with a larger difference between the spectral types of the two stars (hence an increased wind contrast). This leads to a natural experiment probing collision properties with (mostly) one parameter varied at a time. Note that this paper examines the X-ray emissions and a companion paper will focus on the spectropolarimetric behaviour of the systems, to draw a complete picture of the colliding winds in these short-period systems (Johnson et al., in prep, see \citealt{jon19} for preliminary results). Section 2 presents the data used for this analysis which is then explained in Sections 3 and 4 for WR21 and WR31, respectively. Section 5 discusses the results and Section 6 summarizes them.

\section{Data}
\subsection{X-ray domain}

The targets were previously observed by \xmm\ in 2016--20, and we complement these data with new monitoring campaigns in 2021--22 (PI: Naz\'e). A summary is provided in Table \ref{donx}. Both old and new \xmm\ data were processed with the Science Analysis Software (SAS) v20.0.0 using calibration files available in mid-February 2022. The usual pipeline processing was first applied to the European Photon Imaging Camera (EPIC) data. A filtering was then done to keep only the best-quality data ({\sc{pattern}} 0--12 for MOS and 0--4 for pn). We then checked for the presence of background flares by building light curves for energies above 10\,keV, where astrophysical sources emit few photons. Whenever flares were detected, the contaminated intervals were discarded using thresholds on lightcurve count rates. The exposure with ObsID 0880000801 was badly affected by such background flares, and exposure even had to be stopped: only 2\,ks are left after background cleaning, leading to low-quality results. Fortunately, that observation was repeated later, to cover the same orbital phase with a higher signal-to-noise ratio (SNR). The exposure with ObsID 0861710201 displays a high background level, but no specific flare could be identified, so no cleaning was done. For WR\,21, a thick filter was used to reject optical light in all but one exposure (ObsID=0860650501) that rather used medium filter: this has no impact on the derivation of X-ray properties as the results from this exposure are similar to those of the following one (ObsID=0880000301), taken at the same orbital phase (see below). For WR\,31, a thin filter was used to reject optical light in all MOS2 exposures while for pn and MOS1 cameras, a thin filter was used for the first two exposures and a medium one for the others but, again, this has no impact as the results from all cameras agree with each other (see below).

Source detection algorithms (task {\sc edetect\_chain}) were then applied to images binned by a factor of 20 (1\arcsec\ pixels), using likelihoods of 8--10, in order to get count rates in the total (0.5--10.\,keV), soft (0.5--2.\,keV), and hard (2.--10.\,keV) energy bands (Table \ref{donx}). Note that for ObsID 0860650501, WR21 appears off-axis and falls onto a dead CCD of the MOS1 camera: there is therefore no MOS1 data available for that exposure. X-ray spectra and their calibration matrices were then built for each exposure, using the same extraction regions for all datasets of the same target. The source regions were circles of 30\arcsec\ radius centered on the Simbad positions, while background regions were taken in nearby circles of 40--50\arcsec\ radius and devoid of sources. It must be noted that, in all exposures, the background appears uniform, without any problem due to straylight or diffuse emission (see appendix). A binning was applied to all \xmm\ spectra to obtain an oversampling factor of a maximum of five and a minimum signal-to-noise ratio of 3. Light curves were extracted in the same regions as the spectra, in the same energy bands as for the detection run, and for 1\,ks time bins. They were then processed using {\sc epiclccorr} to correct for vignetting, reduced sensitivity off-axis, and bad pixels. We further discarded the noisy bins having effective exposure times lower than 50\% of the time bin length.

All X-ray spectra were then fitted using {\sc Xspec} v12.11.1 and adopting solar abundances from \citet{asp09}. The considered models were absorbed optically thin thermal emissions, e.g. $phabs(ISM) \times phabs \times apec$. The interstellar absorption was evaluated from the known reddenings of the targets (see \citealt{nazwr} for details). Specific details on the spectral fitting are provided below for each target. 

\begin{figure}
  \begin{center}
    \includegraphics[width=9cm]{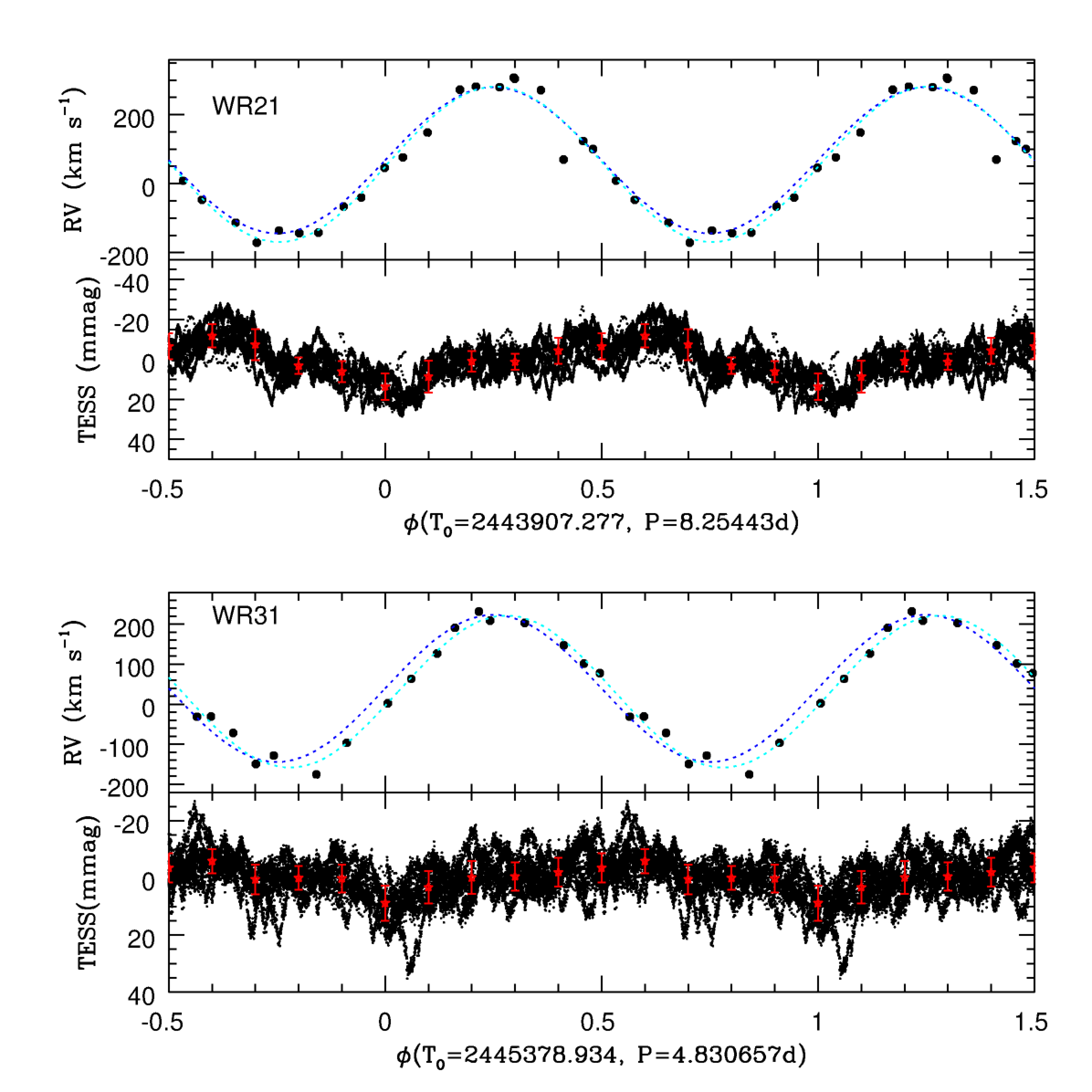}
  \end{center}
  \caption{Radial velocities of the WN components and {\it TESS} photometry of WR21 (top) and WR31 (bottom) phased with the literature ephemeris. The dark blue dotted lines provide the literature orbital solution while the cyan lines yield the new best-fit orbital solution (see text for details). The binned photometry is shown with red stars.}
\label{ephem}
\end{figure}

\begin{table*}
  \scriptsize
  \caption{Observed X-ray characteristics of the targets.  \label{donx}}
  \begin{tabular}{llcccccccccccc}
    \hline
Date &ObsID & HJD & Exp. & $\phi$ & \multicolumn{9}{c}{Net count rates ($10^{-2}$\,cts\,s$^{-1}$)}  \\
&      & $-2.45$& (ks) &     & \multicolumn{3}{c}{MOS1} & \multicolumn{3}{c}{MOS2} & \multicolumn{3}{c}{pn}  \\
&      & $\times 10^6$  &      &     & tot & soft & hard & tot & soft & hard & tot & soft & hard\\
\hline                                               
\multicolumn{12}{l}{\it WR21}\\
15 Aug. 2016&0780070701 &7616.360 & 4.4& 0.81 &2.33$\pm$0.27& 0.97$\pm$0.17& 1.33$\pm$0.20& 2.43$\pm$0.24& 1.19$\pm$0.16& 1.24$\pm$0.17& 6.31$\pm$0.46& 3.13$\pm$0.32& 3.09$\pm$0.33\\
24 Jan. 2021&0860650501 &9238.957 &34.9& 0.38 &             &              &              & 2.47$\pm$0.13& 1.28$\pm$0.09& 1.31$\pm$0.11& 6.70$\pm$0.25& 4.12$\pm$0.19& 2.70$\pm$0.17\\
30 June 2021&0880000301 &9395.876 &11.9& 0.39 &2.46$\pm$0.16& 1.52$\pm$0.13& 0.92$\pm$0.10& 2.41$\pm$0.15& 1.38$\pm$0.11& 1.02$\pm$0.10& 7.17$\pm$0.30& 4.14$\pm$0.22& 2.93$\pm$0.19\\
01 July 2021&0880000401 &9397.455 &22.5& 0.59 &4.50$\pm$0.17& 2.81$\pm$0.13& 1.63$\pm$0.10& 4.64$\pm$0.16& 3.07$\pm$0.13& 1.51$\pm$0.09& 13.2$\pm$0.29& 8.97$\pm$0.24& 3.97$\pm$0.16\\
03 Aug. 2021&0880000501 &9429.748 &16.2& 0.50 &2.66$\pm$0.15& 1.62$\pm$0.12& 1.01$\pm$0.09& 2.55$\pm$0.14& 1.60$\pm$0.11& 0.93$\pm$0.08& 7.47$\pm$0.26& 4.67$\pm$0.20& 2.71$\pm$0.16\\
07 Aug. 2021&0880000601 &9433.921 &18.0& 0.00 &1.70$\pm$0.12& 0.60$\pm$0.07& 1.08$\pm$0.09& 1.88$\pm$0.11& 0.88$\pm$0.08& 0.98$\pm$0.08& 5.00$\pm$0.21& 1.96$\pm$0.13& 3.01$\pm$0.16\\
\multicolumn{12}{l}{\it WR31}\\
26 Dec. 2019&0840210401 &8844.259 & 4.1& 0.45 &2.50$\pm$0.23 &1.18$\pm$0.16 &1.29$\pm$0.17 &2.47$\pm$0.21 &1.05$\pm$0.14 &1.40$\pm$0.16 &6.43$\pm$0.49 &2.40$\pm$0.30 &3.97$\pm$0.38\\
13 Dec. 2020&0861710201 &9197.219 &17.2& 0.51 &2.56$\pm$0.17 &1.62$\pm$0.13 &0.89$\pm$0.11 &2.60$\pm$0.14 &1.25$\pm$0.10 &1.35$\pm$0.10 &7.28$\pm$0.32 &3.54$\pm$0.21 &3.60$\pm$0.24\\
22 Dec. 2021&0880001001 &9571.459 & 7.7& 0.98 &0.93$\pm$0.12 &0.28$\pm$0.07 &0.65$\pm$0.10 &1.27$\pm$0.14 &0.41$\pm$0.08 &0.85$\pm$0.11 &3.30$\pm$0.26 &1.17$\pm$0.16 &2.11$\pm$0.21\\
26 Dec. 2021&0880000901 &9575.035 &10.1& 0.72 &2.21$\pm$0.17 &0.91$\pm$0.11 &1.29$\pm$0.13 &2.35$\pm$0.16 &0.93$\pm$0.10 &1.41$\pm$0.13 &6.54$\pm$0.31 &3.03$\pm$0.21 &3.45$\pm$0.23\\
03 Jan. 2022&0880000801 &9583.082 & 2.4& 0.39 &2.67$\pm$0.51 &1.08$\pm$0.31 &1.57$\pm$0.39 &3.63$\pm$0.69 &2.05$\pm$0.51 &1.58$\pm$0.46 &9.42$\pm$0.86 &4.20$\pm$0.55 &5.16$\pm$0.65\\
10 Jan. 2022&0880000701 &9589.589 & 9.1& 0.74 &2.19$\pm$0.17 &0.91$\pm$0.11 &1.26$\pm$0.13 &2.02$\pm$0.16 &0.69$\pm$0.09 &1.32$\pm$0.13 &6.47$\pm$0.33 &2.57$\pm$0.20 &3.86$\pm$0.25\\
23 Jan. 2022&0880001201 &9603.407 &13.5& 0.60 &2.74$\pm$0.16 &1.35$\pm$0.11 &1.35$\pm$0.11 &2.89$\pm$0.16 &1.53$\pm$0.11 &1.33$\pm$0.11 &7.98$\pm$0.30 &4.14$\pm$0.21 &3.74$\pm$0.20\\
20 Feb. 2022&0880001301 &9631.404 & 5.3& 0.39 &3.04$\pm$0.23 &1.24$\pm$0.15 &1.76$\pm$0.18 &2.65$\pm$0.21 &1.22$\pm$0.14 &1.41$\pm$0.15 &8.57$\pm$0.50 &4.10$\pm$0.34 &4.38$\pm$0.35\\
\hline
  \end{tabular}
  
{\scriptsize For \xmm, exposure times are provided for EPIC-pn data after filtering for flares. For WR21, phases were derived using $P=8.25443$\,d and $T_0=2\,458\,096.680$. For WR31, phases were derived using $P=4.830657$\,d and $T_0=2\,459\,330.004$. In all cases, orbits are circular and phase zero corresponds to the WR being in front ($\phi=0.5$ to the O star in front). The total, soft, and hard energy bands correspond to 0.5--10.0\,keV, 0.5--2.0\,keV, and 2.0--10.0\,keV, respectively. }
\end{table*}

\subsection{\it TESS}

WR21 and WR31 were first observed by {\it TESS} in Sectors 10 and 10+11, respectively \citep{nazwr}. To create the light curves, 50$\times$50 pixel cutouts of the full frame images (FFI) with 30\,min cadence were analyzed with {\sc Lightkurve}\footnote{https://docs.lightkurve.org/}. Source regions were defined by using a large threshold (5--10 times the median absolute deviation above the median flux) while background regions were chosen as pixels with fluxes below the median flux. Two background evaluations were attempted, using a principal component analysis (PCA, with five components) or a simple median. The PCA subtraction appeared slightly better (less long-term trends) hence was used for all light curves. New observations were acquired subsequently, with a 2\,minute cadence, in Sectors 36+37 for WR21 and 37 for WR31. The light curves were downloaded from MAST archives\footnote{https://mast.stsci.edu/portal/Mashup/Clients/Mast/Portal.html} and filtered to keep only the corrected points of highest quality (null quality flag). In all cases, fluxes were converted into magnitudes and the average value was subtracted.

\subsection{Ephemeris check}

The existing ephemerides for the two systems rely on data taken several decades ago, with $T_0<2\,450\,000$ \citep{fah}. In view of the short periods of the systems, a check needs to be done to ensure a correct interpretation of the data.

A series of period search algorithms \citep{aov,hmm,gra13} was applied to the {\it TESS} photometry of each system combining all available sector data. The results agree with those reported in \citet{nazwr}. For WR21 and WR31, the orbital periods show up very clearly in the periodograms: the derived values are 8.354$\pm$0.009\,d and 4.864$\pm$0.003\,d, respectively. Those values are in agreement with the known orbital periods, but their errors are larger because of the shorter temporal baseline. We therefore keep the period values from literature for this paper. Once folded, the {\it TESS} photometry reveals shallow atmospheric eclipses when the WR star is in front for both systems (Fig. \ref{ephem}). The broadness and shallowness of the feature, combined with the higher-frequency photometric variability of the stars, makes it difficult to pinpoint a very precise $T_0$, however. Radial velocities (RVs) are more useful for this purpose.

Previous orbital solutions were made by measuring RVs on the N\,{\sc v}\,$\lambda$4603.2\,\AA\ line (same rest wavelength as \citealt{fah}), which is linked to the WR wind. This line is clearly detected in the flux spectra of our spectropolarimetric data (obtained with the RSS instrument on the Southern African Large Telescope, see \citealt{kob03,buc06} in general and Johnson et al., in prep, specifically for this dataset). The resolution of these spectra is quite low (pixel size = 1\,\AA) but, fortunately, the amplitude of the orbital variations is large. We determined the position of the line maximum by fitting a Gaussian to the top 5 pixels of this emission line (Table \ref{rv}) and then fitted the derived RVs with a sine wave using a least square algorithm. An error of 20\,\kms\ was adopted for individual RVs.

\begin{figure*}
  \begin{center}
    \includegraphics[width=8.5cm,bb=20 516 580 690, clip]{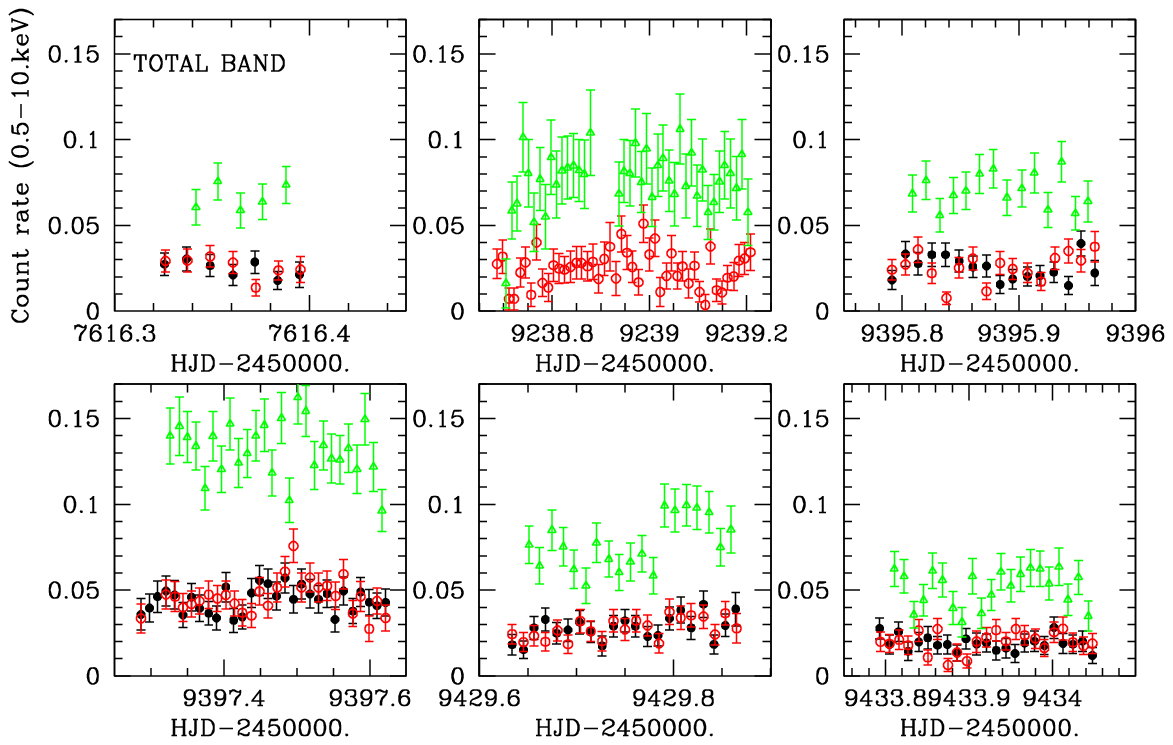}
    \includegraphics[width=8.5cm,bb=20 340 580 516, clip]{WR21_lc_tot.ps}
    \includegraphics[width=8.5cm,bb=20 516 580 690, clip]{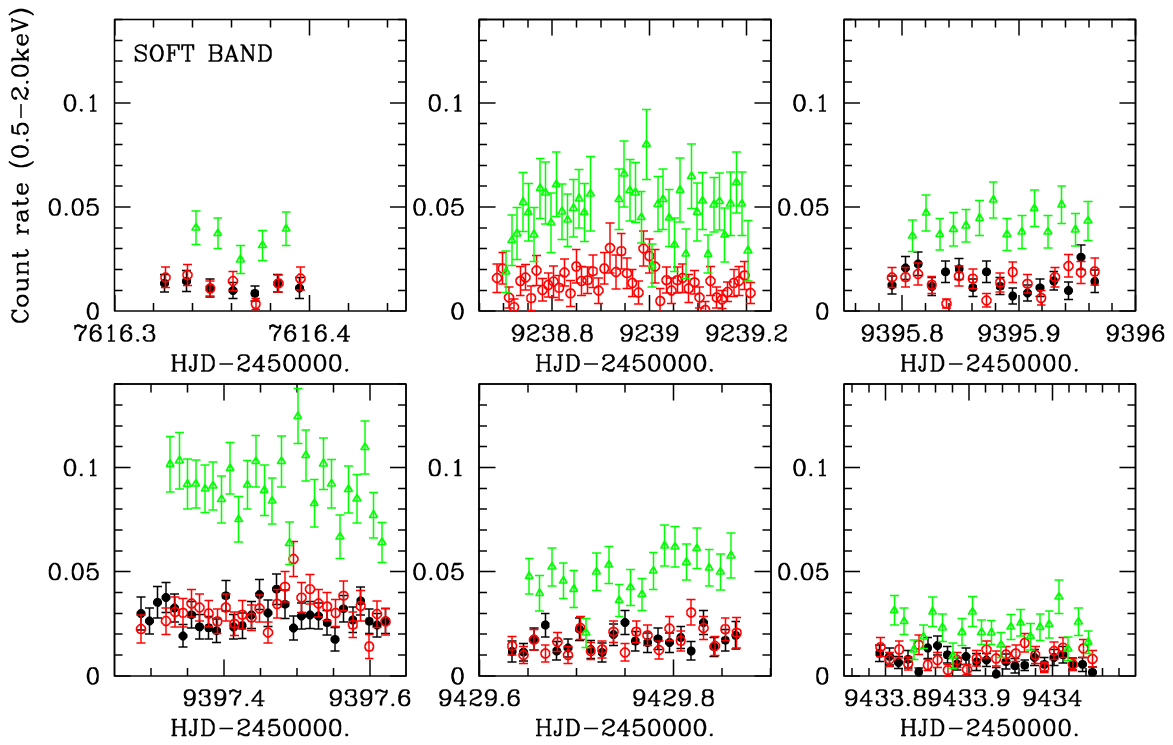}
    \includegraphics[width=8.5cm,bb=20 340 580 516, clip]{WR21_lc_soft.ps}
    \includegraphics[width=8.5cm,bb=20 516 580 690, clip]{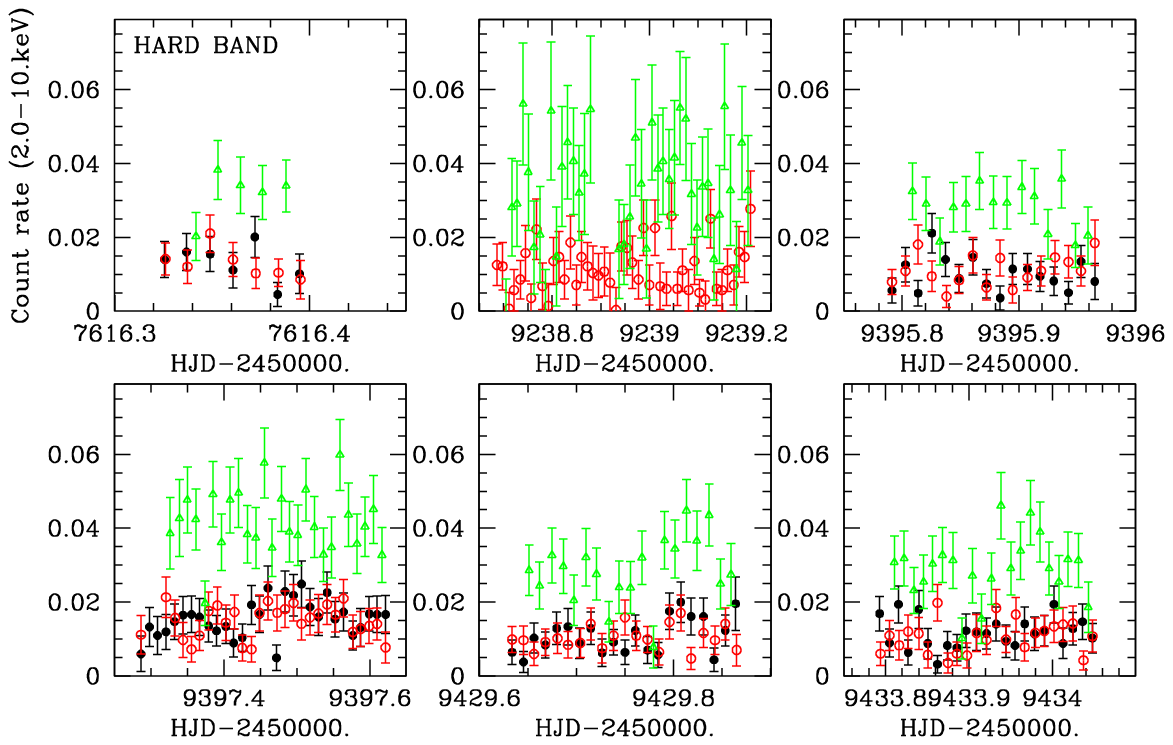}
    \includegraphics[width=8.5cm,bb=20 340 580 516, clip]{WR21_lc_hard.ps}
  \end{center}
  \caption{The X-ray light curves in each of the six exposures of WR21, with the top, middle, and bottom panels showing the total, soft and hard band data, respectively. Green symbols correspond to EPIC-pn, black to EPIC-MOS1, and red to EPIC-MOS2.}
\label{wr21lc}
\end{figure*}

For WR21, the best-fit amplitude is $K=225\pm6$\,km\,s$^{-1}$, slightly larger than the $K=211.8\pm2.5$\,km\,s$^{-1}$ of \citet{fah}, but in good agreement considering errors (our rms is 34\,km\,s$^{-1}$ - the rms associated to the literature solution is 36\,km\,s$^{-1}$). The best-fit solution also indicates a time of conjunction with the WR star in front $T_0$ of 2\,458\,096.680$\pm$0.037, while the combined {\it TESS} photometry suggests a $T_0$ of 2\,459\,327.169 (using only the oldest {\it TESS} sectors yields 2\,458\,591.912, see \citealt{nazwr}). For all these $T_0$ values, the phase shifts with respect to \citet{fah} ephemeris are always small ($\Delta\phi<0.075$, see top panel of Fig. \ref{ephem}). Because the {\it TESS} photometry seems to suffer from the stochastic stellar variations and the orbital variations are better defined in the RVs due to their large amplitude, we adopt the $T_0=2\,458\,096.680$ from the RV solution in this paper (with $\phi=0$ corresponding to the WR being in front of its companion).

For WR31, the best-fit RV amplitude amounts to $K=190\pm7$\,km\,s$^{-1}$, again in agreement with the previously reported one ($K=183\pm3$\,km\,s$^{-1}$, \citealt{fah}) considering errors. The rms value is 18\,km\,s$^{-1}$ for our new solution, to be compared to an rms of 31\,km\,s$^{-1}$ when using the literature solution. Both best-fit values of $T_0$ from the combined photometry and the radial velocities correspond to a small shift, $\Delta\phi=+0.03$, with respect to the ephemeris of \citet{fah} - see bottom panel of Fig. \ref{ephem}. Using only the oldest {\it TESS} sectors already yielded $\Delta\phi=+0.02$, see \citet{nazwr}. In this paper, we then adopt the more recent photometric $T_0=2\,459\,330.004$ (with $\phi=0$ corresponding to the WR being in front of its companion).

\begin{table}
  \scriptsize
    \caption{Radial velocities of WR21 and WR31 ($\sigma=20$\,km\,s$^{-1}$).}
\label{rv}
\begin{tabular}{lclc}
  \hline
  \multicolumn{2}{c}{WR21} & \multicolumn{2}{c}{WR31} \\
HJD-2450000. & RV (\kms) & HJD-2450000. & RV (\kms) \\
         \hline
8090.571 &  279.5 & 8840.533 &  --149.2 \\
8135.466 &--171.0 & 8843.533 &    203.1 \\
8152.396 &--135.7 & 8862.469 &    209.0 \\
8173.385 &  307.4 & 8863.517 &    101.9 \\
8174.333 &   69.6 & 9202.579 &   --71.9 \\ 
8177.545 &--143.5 & 9213.508 &   --96.3 \\ 
8238.390 &  272.5 & 9214.516 &    126.6 \\
8815.584 &  148.3 & 9219.544 &    190.9 \\ 
8822.573 & --40.2 & 9221.497 &   --31.1 \\
8839.529 &   46.3 & 9267.366 &     63.4 \\
8842.509 &  271.1 & 9295.290 &  --175.1 \\
8843.503 &  100.1 & 9303.283 &     77.9 \\
8846.523 &--141.6 & 9304.475 &  --128.0 \\
8849.525 &  281.3 & 9629.402 &      2.3 \\
8860.448 &    8.6 & 9630.419 &    231.7 \\
8861.451 &--113.0 & 9655.520 &    147.1 \\
9186.564 &   76.2 & 9714.381 &   --30.4 \\ 
9206.509 &  123.4  \\
9207.494 & --46.8  \\
9309.250 & --66.1  \\
9758.253 &  304.4  \\
        \hline
\end{tabular}
\end{table}

\begin{figure*}
  \begin{center}
    \includegraphics[width=8.5cm]{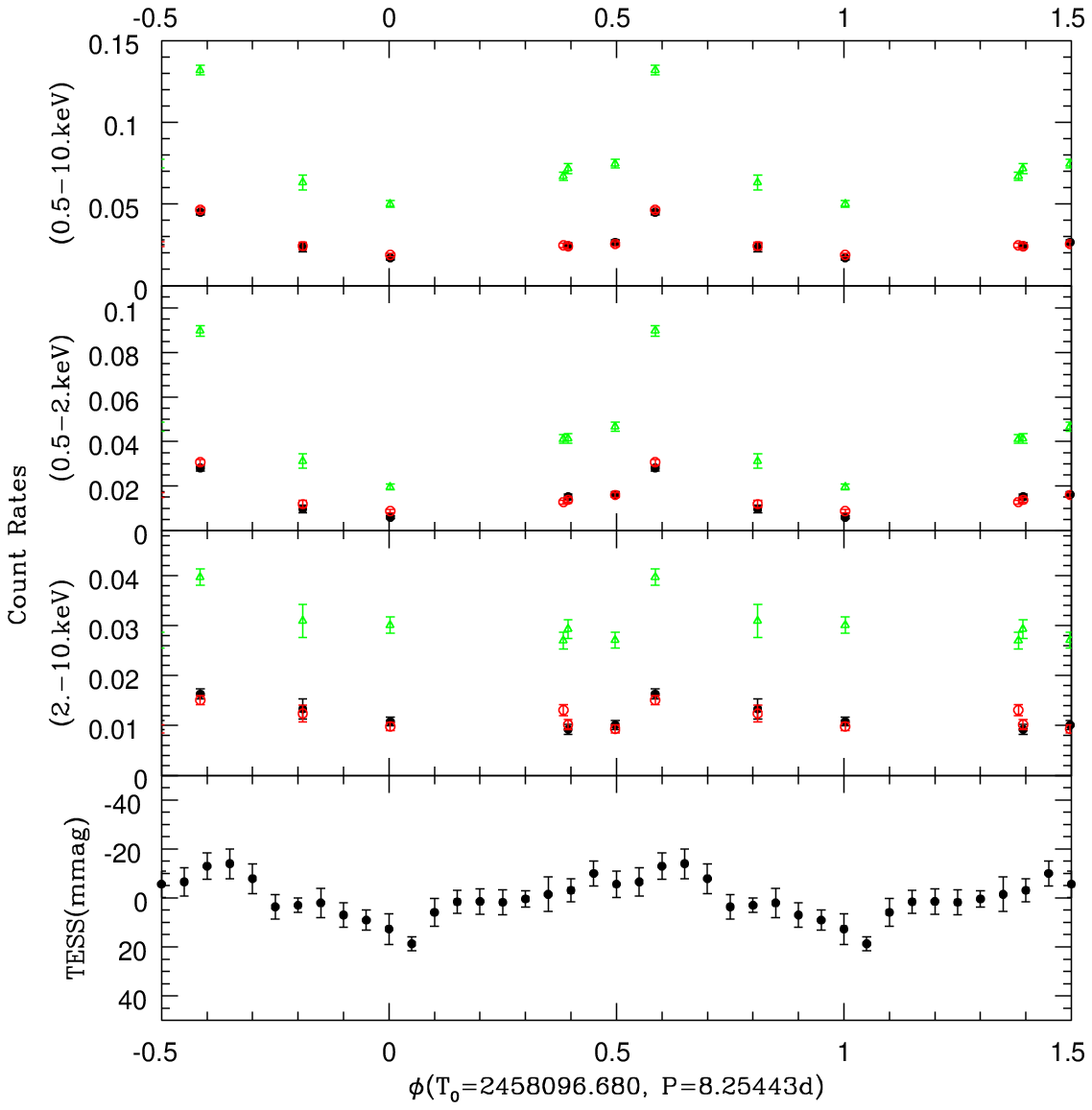}
    \includegraphics[width=8.5cm]{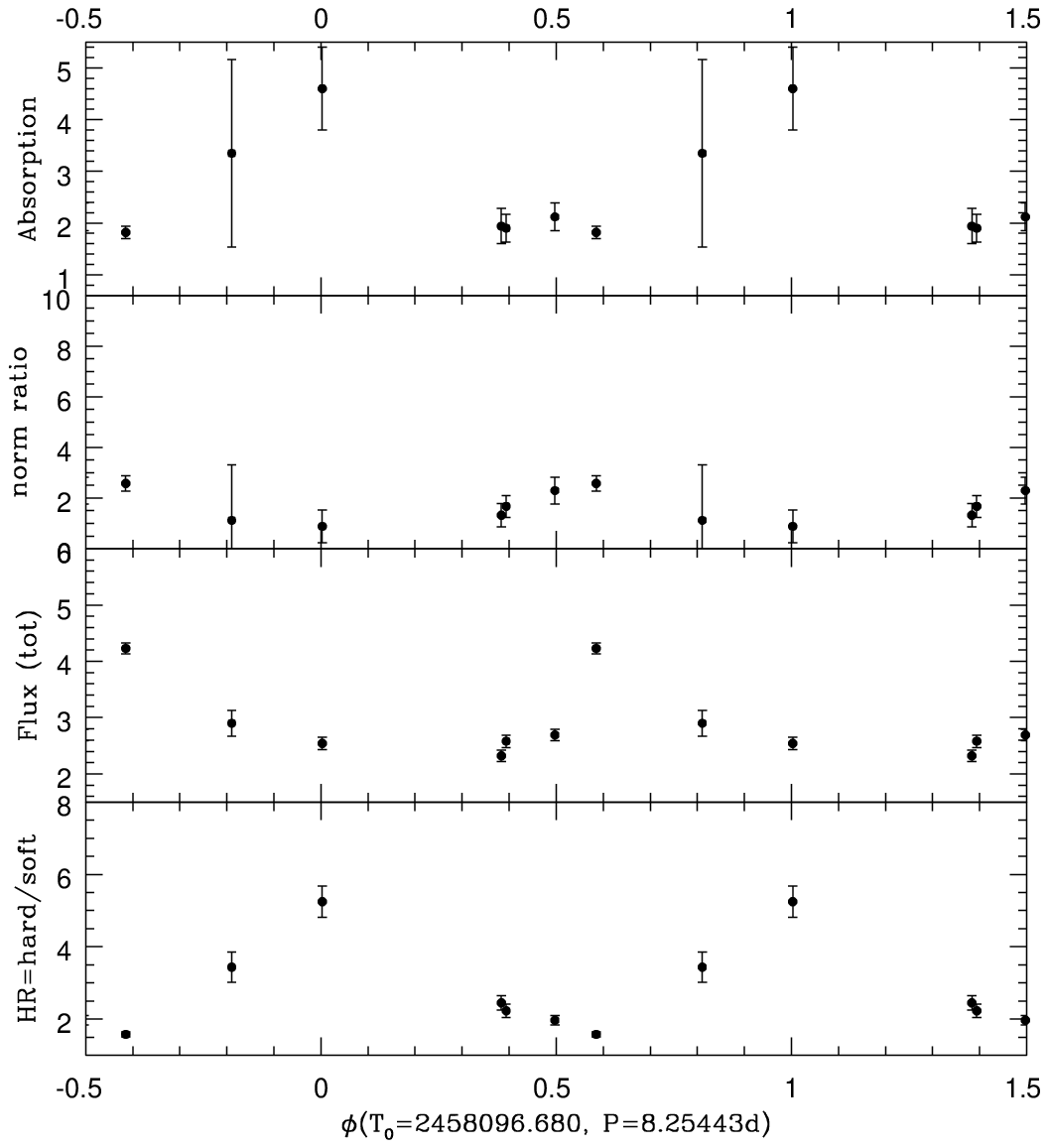}
  \end{center}
  \caption{Left: \xmm\ count rates of WR21 in the three energy bands as a function of phase (see Sect. 2.3), compared to {\it TESS} average photometry (equivalent of red stars in Fig. \ref{ephem}). EPIC-pn, MOS1, MOS2 data are shown with green triangles, black dots, and red open squares, respectively. Right: Results from the spectral fitting using a model $phabs(ISM)\times [ phabs \times apec(0.3) + phabs \times \{apec(0.9)+apec(3.0) \}]$ with solar abundances. From top to bottom are shown the absorption of the warmest thermal components (i.e. 0.9 and 3.0\,keV), the ratio of their normalization factors $norm(0.9)/norm(3.0)$, the observed flux in total band, and the ratio $HR$ between the hard observed X-ray flux and the soft one. }
\label{wr21cr}
\end{figure*}

\begin{table*}
  \scriptsize
  \caption{Best-fit models to the X-ray spectra of WR21. \label{wr21fit}}
  \begin{tabular}{lcccccccccccc}
    \hline
\multicolumn{13}{l}{Solar abundances, $phabs(ISM)\times [ phabs \times apec(0.3) + phabs \times \{apec(0.9)+apec(3.0)\}]$, first two columns fixed to $0.41\times10^{22}$ and $10^{22}$\,cm$^{-2}$}\\
ObsID & $\phi$ & $norm$(0.3\,keV) & $N_{\rm H}$ & $norm$(0.9\,keV) & $norm$(3.0\,keV) & $\chi^2$/dof & \multicolumn{3}{c}{$F_{\rm X}^{\rm obs}$($10^{-13}$\,erg\,cm$^{-2}$\,s$^{-1}$)}  & \multicolumn{3}{c}{$F_{\rm X}^{\rm ISM-cor}$($10^{-13}$\,erg\,cm$^{-2}$\,s$^{-1}$)}\\
     & & ($10^{-4}$\,cm$^{-5}$) &($10^{22}$\,cm$^{-2}$) & ($10^{-4}$\,cm$^{-5}$) & ($10^{-4}$\,cm$^{-5}$) & & tot & soft & hard& tot & soft & hard\\
\hline
0780070701 &0.81 &7.72$\pm$1.64 &3.35$\pm$1.81 &4.25$\pm$8.29 &3.79$\pm$0.71 &34.54/42   &2.90$\pm$0.23 &0.66$\pm$0.06 &2.25$\pm$0.20 & 3.41$\pm$0.27 & 1.08$\pm$0.09 & 2.32$\pm$0.21 \\
0860650501 &0.38 &3.70$\pm$1.38 &1.94$\pm$0.34 &3.19$\pm$1.06 &2.41$\pm$0.23 &75.70/94   &2.32$\pm$0.10 &0.67$\pm$0.03 &1.65$\pm$0.11 & 2.74$\pm$0.12 & 1.02$\pm$0.05 & 1.71$\pm$0.11 \\
0880000301 &0.39 &3.80$\pm$1.29 &1.90$\pm$0.27 &4.13$\pm$0.99 &2.47$\pm$0.25 &79.57/105  &2.58$\pm$0.11 &0.80$\pm$0.04 &1.78$\pm$0.12 & 3.05$\pm$0.13 & 1.20$\pm$0.06 & 1.85$\pm$0.12 \\
0880000401 &0.59 &10.80$\pm$1.16&1.82$\pm$0.12 &8.19$\pm$0.79 &3.18$\pm$0.21 &218.93/240 &4.23$\pm$0.10 &1.64$\pm$0.04 &2.59$\pm$0.10 & 5.28$\pm$0.12 & 2.57$\pm$0.06 & 2.70$\pm$0.10 \\
0880000501 &0.50 &6.18$\pm$1.17 &2.12$\pm$0.27 &5.35$\pm$1.11 &2.33$\pm$0.23 &139.96/143 &2.69$\pm$0.10 &0.91$\pm$0.03 &1.79$\pm$0.10 & 3.26$\pm$0.12 & 1.40$\pm$0.05 & 1.86$\pm$0.10 \\
0880000601 &0.00 &5.57$\pm$0.43 &4.60$\pm$0.80 &3.61$\pm$2.66 &4.08$\pm$0.27 &138.96/115 &2.54$\pm$0.11 &0.41$\pm$0.03 &2.13$\pm$0.10 & 2.89$\pm$0.13 & 0.69$\pm$0.05 & 2.19$\pm$0.10 \\
\hline
\multicolumn{13}{l}{WNE abundances, $phabs(ISM)\times [ vphabs \times vapec(0.3) + vphabs \times \{vapec(0.9)+vapec(3.0)\}]$, first two columns fixed to $0.41\times10^{22}$ and $1.72\times 10^{20}$\,cm$^{-2}$}\\
ObsID & $\phi$ & $norm$(0.3\,keV) & $N_{\rm H}$ & $norm$(0.9\,keV) & $norm$(3.0\,keV) & $\chi^2$/dof & \multicolumn{3}{c}{$F_{\rm X}^{\rm obs}$($10^{-13}$\,erg\,cm$^{-2}$\,s$^{-1}$)} & \multicolumn{3}{c}{$F_{\rm X}^{\rm ISM-cor}$($10^{-13}$\,erg\,cm$^{-2}$\,s$^{-1}$)} \\
     & & ($10^{-5}$\,cm$^{-5}$) &($10^{20}$\,cm$^{-2}$) & ($10^{-5}$\,cm$^{-5}$) & ($10^{-5}$\,cm$^{-5}$) & & tot & soft & hard& tot & soft & hard\\
\hline
0780070701 &0.81 & 1.81$\pm$0.35 &6.59$\pm$3.13 &0.76$\pm$1.40 &1.31$\pm$0.18 & 34.03/42  &2.90$\pm$0.22 &0.66$\pm$0.06 &2.25$\pm$0.21 & 3.40$\pm$0.26 & 1.07$\pm$0.10 & 2.32$\pm$0.22 \\
0860650501 &0.38 & 0.99$\pm$0.29 &4.50$\pm$0.82 &0.99$\pm$0.32 &0.81$\pm$0.08 & 75.76/94  &2.31$\pm$0.12 &0.68$\pm$0.04 &1.63$\pm$0.11 & 2.72$\pm$0.14 & 1.02$\pm$0.05 & 1.70$\pm$0.11 \\
0880000301 &0.39 & 0.96$\pm$0.27 &4.20$\pm$0.61 &1.22$\pm$0.29 &0.85$\pm$0.09 & 81.42/105 &2.57$\pm$0.11 &0.80$\pm$0.04 &1.77$\pm$0.11 & 3.03$\pm$0.13 & 1.19$\pm$0.06 & 1.84$\pm$0.11 \\
0880000401 &0.59 & 2.68$\pm$0.26 &4.09$\pm$0.26 &2.55$\pm$0.24 &1.08$\pm$0.07 &218.33/240 &4.21$\pm$0.10 &1.64$\pm$0.04 &2.60$\pm$0.12 & 5.25$\pm$0.12 & 2.56$\pm$0.06 & 2.69$\pm$0.12 \\
0880000501 &0.50 & 1.51$\pm$0.25 &4.69$\pm$0.58 &1.58$\pm$0.31 &0.80$\pm$0.08 &139.43/143 &2.69$\pm$0.10 &0.91$\pm$0.04 &1.78$\pm$0.10 & 3.25$\pm$0.12 & 1.40$\pm$0.05 & 1.85$\pm$0.10 \\
0880000601 &0.00 & 1.33$\pm$0.10 &10.01$\pm$1.77&0.71$\pm$0.62 &1.36$\pm$0.09 &142.77/115 &2.51$\pm$0.10 &0.40$\pm$0.02 &2.10$\pm$0.10 & 2.85$\pm$0.11 & 0.68$\pm$0.04 & 2.17$\pm$0.10 \\
\hline
  \end{tabular}
  
{\scriptsize Total, soft and hard energy bands being defined as 0.5--10.0\,keV, 0.5--2.0\,keV, and 2.0--10.0\,keV, respectively. Errors are 1$\sigma$ uncertainties; they correspond to the larger value if the error bar is asymmetric. WNE abundances correspond to He, C, N, O and other elements fixed to 100, 1, 550, 4.4, and 28 times solar, respectively \citep{nazwr}. Note that using the tbabs model for the interstellar absorption yields very similar results.}
\end{table*}

\section{WR21 (WN5+O7V, $P=8.25443$\,d)}
\subsection{X-ray light curves and spectra}
The light curves of WR\,21 in individual exposures do not show significant variations (Fig. \ref{wr21lc}). In particular, they do not reveal any obvious signature of eclipses although the errors are large. In contrast, differences between exposures are obvious (Fig. \ref{wr21cr}), with a minimum near $\phi=0$ and an increase peaking shortly before $\phi=0.6$. Such changes are particularly strong in the soft energy band, with a factor of 4 increase in count rate between minimum and maximum while the increase is reduced to 50\% for the hard band (Table \ref{donx}). This suggests a leading role of absorption as it affects mostly soft X-rays. Interestingly, the maximum is not seen at $\phi=0.5$ but at $\phi=0.6$, as in V444\,Cyg \citep{lom15}. This delayed peak is probably due to Coriolis deflection of the shock cone. The half-amplitude peak width covers a maximum of 0.2 in phase, suggesting a quite narrow shock cone. Note that the increase in count rate seems quite fast, but its decrease is more difficult to constrain as there is no data point near $\phi=0.7$ so we cannot securely assess whether the peak truly is symmetric or not. 

We then fit the spectra, considering several spectral models in turn, with the aim to find the simplest model able to reproduce the spectra of all exposures. That implies that the highest quality spectra will be the most constraining ones and that there might sometimes be an impression of ``overfitting" for some lower-quality exposures. However, only by applying the same model to all exposures are we able to unveil trends. A single temperature component was clearly not sufficient to reproduce the spectra well. Two emission components with temperatures free to vary and a single local absorption provided better results, but fixing the temperatures of such components resulted in a clear degradation of the residuals. Comparing the spectral shapes between minimum and maximum brightness reveals that the hard ($>$2\,keV) tail remains about the same, as do the very soft ($<$0.5\,keV) parts of the spectra: the changes occur in between those two energies (reminiscent of the case of $\gamma^2$\,Vel, \citealt{rau00}). This suggests that a single absorption may not be the correct way to represent the situation. Therefore, we tried fitting with three emission components of fixed temperatures 0.3, 0.9, and 3.\,keV (to cover all cases found in the unconstrained 2-T fitting) absorbed by two different columns: one for the lowest temperature and one for the two hotter ones. It again yielded good (actually even slightly better) residuals, $\chi_{\nu}^2\sim1$. The absorption of the cool component was not significantly varying hence it was fixed and fits were re-done to reach final results (Table \ref{wr21fit} and Fig. \ref{wr21cr}).

The previous fittings were done using solar abundances for both circumstellar absorption and emissions. Since the material in the collision zone should be a mix of O and WR material, we also considered models with non-solar abundances typical of the WR star, as in \citet{nazwr}. For all components which are not interstellar, these models use abundances relative to Helium inspired from the Galactic WNE models of PoWR\footnote{http://www.astro.physik.uni-potsdam.de/$\sim$wrh/PoWR/powrgrid1.php}. Since abundances in {\sc Xspec} are relative to Hydrogen and relative to solar values, extremely small Hydrogen contents are difficult to simulate. As in \citet{nazwr}, the Helium number abundance relative to Hydrogen was then arbitrarily set to 100 times the solar value (implying a mass fraction of Hydrogen $X$ of 0.028) while values of 1, 550, 4.4, and 28 were used for C, N, O, and other elements, respectively. The best-fit parameters of these models are also listed in Table \ref{wr21fit} and their evolution appears very similar to that of models using solar abundances.

\begin{figure}
  \begin{center}
    \includegraphics[width=8.5cm]{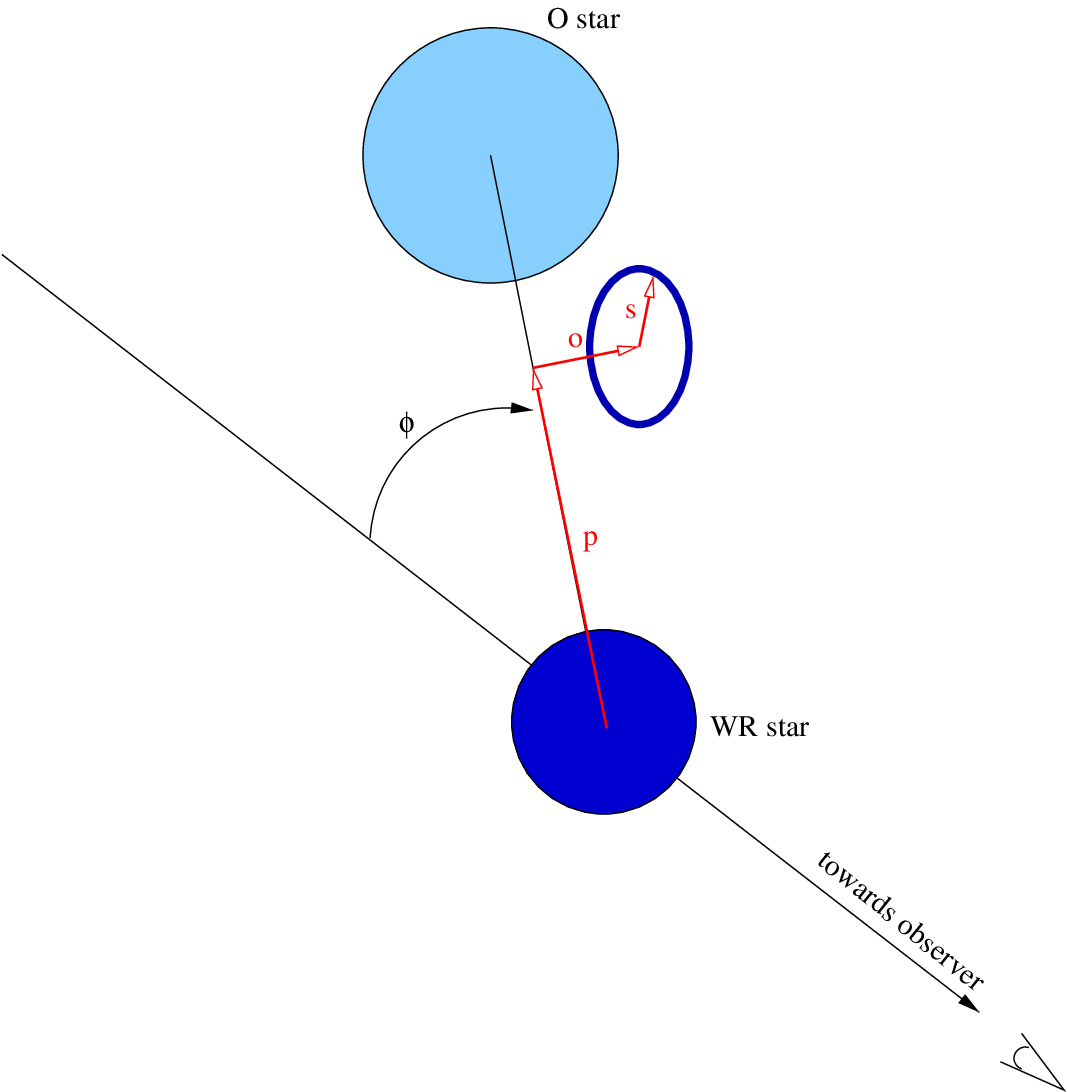}
  \end{center}
  \caption{Sketch of the simple model used in Sections 3.2 and 4.2. }
\label{sketch}
\end{figure}

The cool component with its constant absorption probably arises far in the wind(s). However, the strength of that plasma also appears larger at maximum brightness ($\phi=0.6$): there is probably an additional contribution from the (soft) intrinsic X-ray emission of the O-star wind when it comes into view. The two hotter components suffer from a slightly lower absorption at the same phase and have a larger absorption when the WR star is in front ($\phi=0$) - the absorbing column more than doubles between $\phi=0.6$ and $\phi=0$. The fluxes clearly follow the count rates, with a hardness ratio significantly larger when the WR star is in front of its companion.

\begin{figure*}
  \begin{center}
    \includegraphics[width=8.5cm]{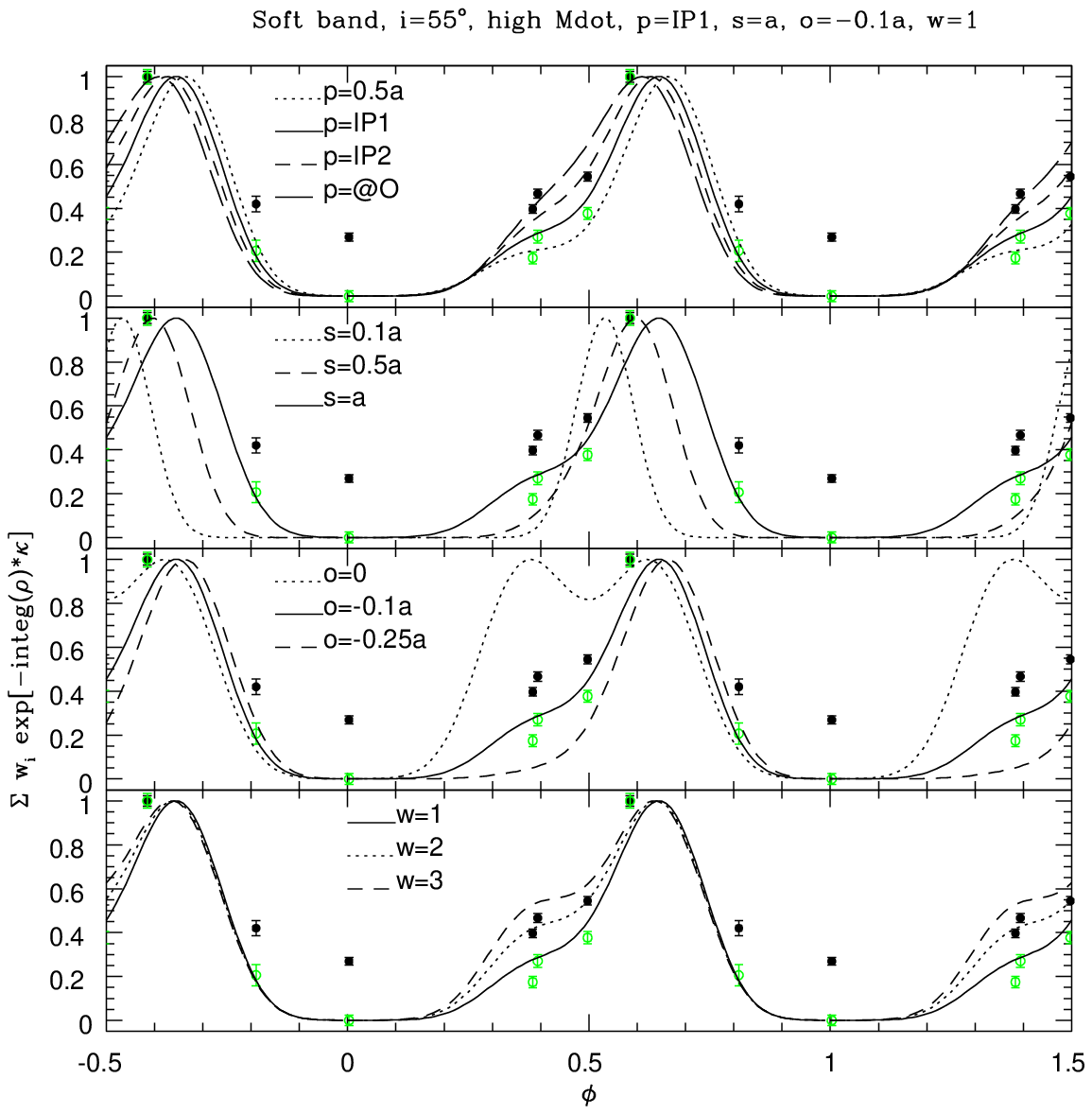}
    \includegraphics[width=8.5cm]{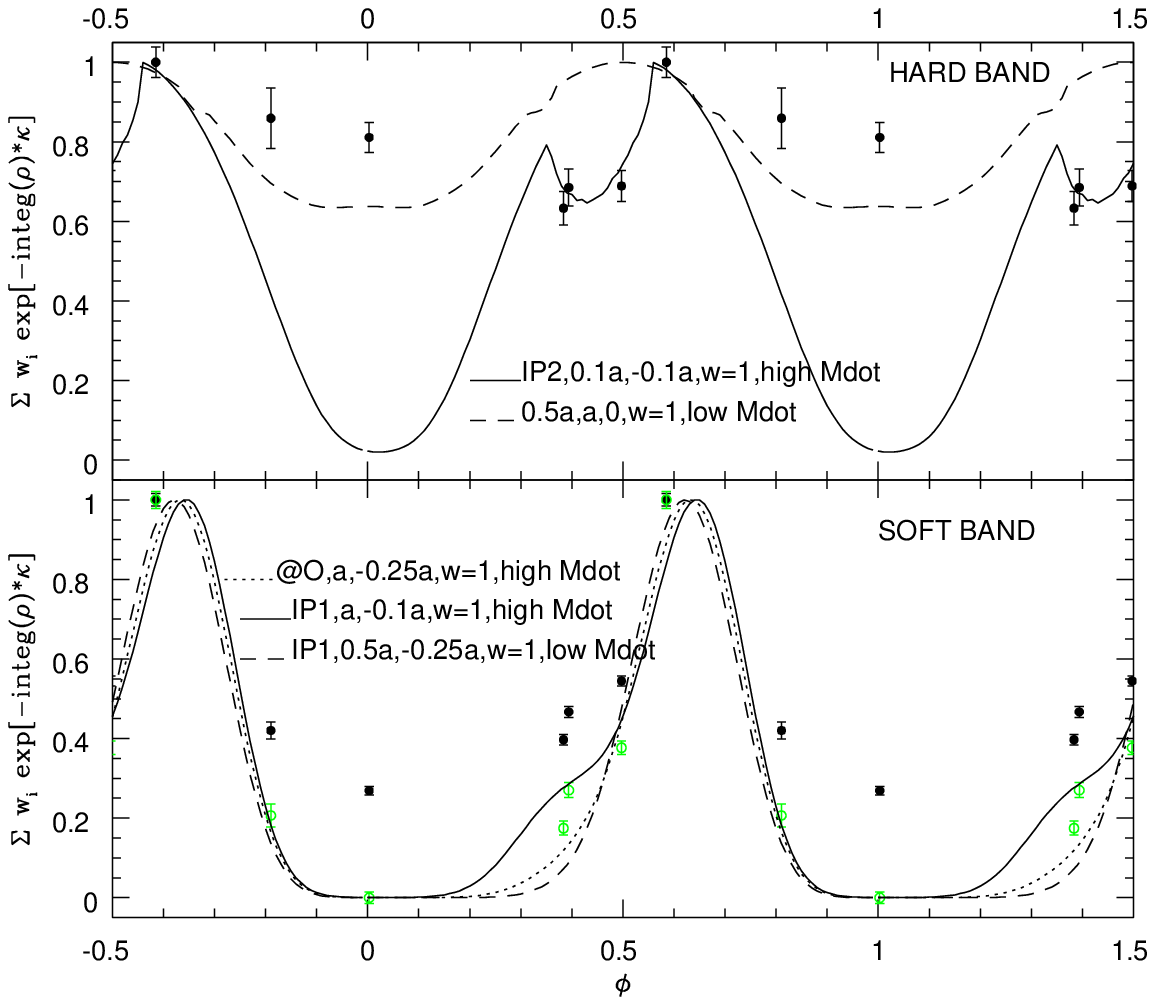}
  \end{center}
  \caption{Left: Predicted soft X-ray lightcurves for the simple model considering wind absorption of colliding wind emission in WR\,21 (Sect. 3.2), normalized by their maximum value, compared to fluxes corrected for ISM absorption in the soft band (Sect. 3.1 and Table \ref{wr21fit}). Black dots correspond to fluxes divided by their maximum value, while green circles provide the same fluxes but with the minimum value subtracted before normalization so that the minimum flux appears at zero and the maximum at one. The model shown with a solid line corresponds to an inclination of 55$^{\circ}$, the higher mass-loss rate, a position in-between the O star and the midpoint between the stars (IP1, see text), an annulus of size equal to the separation ($s=a$), an offset of one tenth of the separation ($o=-0.1a$), and a ring of uniform brightness ($w_{\rm pro}=1$). The dotted and dashed lines in each panel correspond to variations of each parameter in turn. Right: Best models, superimposed to the hard (top) and soft (bottom) fluxes.  }
\label{wr21mod}
\end{figure*}

To assess the importance of the intrinsic wind shocks to the observed X-ray flux, we use the canonical $\log[L_{\rm X}/L_{\rm BOL}]$ relationship. \citet{naz11} provide it for single O stars: excluding outliers, it amounts to $-7.45, -7.84$, and $-9.41$ in the 0.5--1.0\,keV, 1.0--2.5\,keV, and 2.5--10.0\,keV bands, respectively. If the secondary star is an O7V object \citep{fah}, its $\log[L_{\rm BOL}/L_{\odot}]$ should be 5.10 \citep{mar05}. For the WR star, a $\log[L_{\rm BOL}/L_{\odot}]$ of 5.53 is found on average in recent fits of WN5 stars \citep{ham19}. Assuming that both stars follow the canonical relation\footnote{This is clearly optimistic, hence an upper limit, for a WR star.} and adopting a distance of 3230\,pc \citep{bai21}, we derive X-ray fluxes of 7.1 and $0.055\times 10^{-14}$\,erg\,cm$^{-2}$\,s$^{-1}$ for the 0.5--2.5\,keV, and 2.5--10.0\,keV bands, respectively. In contrast, the observed X-ray fluxes, after correction of the interstellar absorption, are $10-33$, and $14-20\times 10^{-14}$\,erg\,cm$^{-2}$\,s$^{-1}$ in the 0.5--2.5\,keV, and 2.5--10.0\,keV bands. It is clear that the intrinsic emission from embedded wind shocks does not contribute significantly to the observed hard flux. For the soft band, the minimum observed flux may be compatible with that expected for embedded wind shocks, although this simple calculation does not take into account the high local absorption by the WR wind and is most probably quite optimistic. 

\subsection{A simple model}

Considering the orbital and wind parameters found by \citet{fah}, there is no pressure equilibrium between the two stellar winds in WR21: the WR wind should crash onto the O star. This however neglects the radiative braking effects: as the separation is small, the wind of one star is accelerated by its own radiation but slowed by the radiation of the companion \citep{gay97}. This results in slower winds and avoids the wind crash. For example, an absence of equilibrium should also occur in V444\,Cyg with the wind parameters taken at face value, but X-ray and polarimetric data clearly showed the presence of a colliding wind region in between the stars \citep{lom15}. A similar situation is expected for WR\,21 as well as for WR\,31.

We can therefore assume that a colliding wind region is present between the stars and emitting X-rays. We must now constrain its size and position, which can be done using the X-ray lightcurve. Contrary to V444\,Cyg, no strong eclipses are detected in the X-ray domain, the flux errors are larger, and the phase coverage of the X-ray light curve remains patchy, prohibiting the derivation of detailed parameters. However, X-ray variations are present and can still broadly be related to a changing viewing angle on the colliding wind region. To this aim, we have built a simple model following what was done for V444\,Cyg (see Sect. 4.1 and  Fig. 12 in \citealt{lom15}, and Fig. \ref{sketch}). In this model, the emission region is modelled as a thin annulus (i.e. a circle), which can be considered as the ``center of gravity'' for the X-ray emission in the energy band under consideration. The modelling, which does not assume any plasma temperature, can be independently applied to several energy bands - in our case, the soft and hard bands defined above. As can be seen on the observed light curves, the soft and hard curves are not identical hence provide different information. Indeed, X-rays of various energies are born at various positions of the collision zone. They also suffer at various degrees of absorption and occultation effects (the former dominating at softer energies, the latter at higher energies). 

In the model, any annulus has a radius $s$, perpendicular to both the orbital plane and the line-of-centers. It is located at a distance $p$ from the WR star along the line-of-centers. To reproduce potential Coriolis effects, the annulus center can be offset by a quantity $o$ from the line-of-centers along the direction of the O-star motion (which is perpendicular to the line-of-centers but in the orbital plane). Negative offset values indicate a point lagging behind the O star. Also linked to orbital motion effects, the leading side is allowed to be brighter than the trailing one by a factor $w_{\rm pro}$. Trials were made using three annulus radii $s$ of 0.1, 0.5, and 1 times the orbital separation $a$, three offsets $o$ of 0, --0.1, and $-0.25\times a$, and three brightness ratios $w_{\rm pro}$ of 1, 2, and 3. Four positions $p$ were considered: $a/2$ (halfway between the stars), $a-1.01\times R_{*,O}$ (close to the O-star surface, labelled ``@O'' in figures), and two equally-spaced intermediate positions between them (labelled IP1 and IP2 in figures, for the one closest to the halfpoint and the one closest to the O star, respectively).

For each annulus, we considered that it is immersed in a spherically symmetric WR wind and then we integrated the wind density $\rho$ from each point of the annulus to the observer, with the line-of-sight determined by the inclination $i$ of the system. Note that our model could overestimate the absorption, as the line-of-sight of some emission regions could pass through the O-star wind at some phases, but that nevertheless provides a good first estimate, even for high-inclination systems for which this effect is larger, as shown for V444\,Cyg in \citet{lom15}. 

The luminosity is then calculated using $\sum w_j e^{-\kappa \int \rho \, dl}$. Occultations by the O star or WR star result in an infinite integral, i.e. the absence of observed emission. In the above equation, $j$ corresponds to a point of the annulus and $w_j$ is its leading vs trailing brightness weighting. The latter is equal to $1+ (w_{\rm pro}-1)\times 0.5\times (\delta/s+1)$ with $s$ the annulus radius, $\delta$ the projected distance of the annulus point onto the line joining the leading edge to the trailing edge ($\delta=+s$ so $w_j=w_{\rm pro}$ at the leading edge point and $\delta=-s$ so $w_j=1$ at the trailing edge point).

The density $\rho$ follows the classical $\dot M/4\pi r^2v(r)$ relationship with the velocity following a $\beta=1$ velocity law (i.e. $v(r)=v_{\infty}\times(1-R_*/r)$). The opacity $\kappa$ of the WR wind was determined from a single {\sc Xspec} absorbed thermal emission model ($phabs(ISM) \times vphabs \times vapec$) using WR abundances (see Sect. 3.1 and \citealt{nazwr}). Fitting the WR21 spectra with such a model yielded on average a temperature of 3\,keV, an equivalent absorbing Hydrogen column of 3 to $7\times 10^{19}$\,cm$^{-2}$ and normalizations of $10^{-5}$\,cm$^{-5}$. As a check, a similar trial was made for WR\,31 spectra and a similar average normalization was found, although with a temperature of 3.6\,keV and absorbing columns of 1 to $3\times 10^{20}$\,cm$^{-2}$. For both stars, the fluxes from these average models, with and without local absorption, were then determined in both soft and hard bands. Since the fluxes suffering absorption are multiplied by $e^{-\tau}$ and $\tau=\kappa \int \rho \, dl=\kappa \, N_{\rm H} \, m_{\rm H}/X$, we derived $\kappa$ of about 5.5 and 100.0\,g\,cm$^{-2}$ for the hard and soft bands, respectively. 

For the stellar parameters of the WR star, we used a stellar radius of 4.95\,R$_{\odot}$, a wind terminal velocity of 1485\,\kms, and a mass-loss rate of $2\times 10^{-5}$\,M$_{\odot}$\,yr$^{-1}$, which are all typical values for WN5 stars from the recent models of \citet{ham19}. A lower mass-loss rate of $10^{-5}$\,M$_{\odot}$\,yr$^{-1}$, advocated by \citet{lam96} for WR\,21, was also tried. The O-star radius was set to 9.37\,R$_{\odot}$, typical of O7V stars \citep{mar05}. Regarding inclination, two values were proposed for the system \citep{fah}: 68.5$^{\circ}$ resulting in a separation $a=57.72$\,R$_{\odot}$ and 49.6$^{\circ}$ resulting in a separation $a=70.52$\,R$_{\odot}$. A new estimate based on modelling polarimetric phase variations leads to 55$\pm$7$^{\circ}$ (Johnson et al., in prep, using the method of \citealt{ful22}), which yields a separation of $a=65.56$\,R$_{\odot}$. All three cases were considered in our models (Table \ref{param}).

\begin{table}
  \scriptsize
    \caption{Stellar and orbital parameters of WR21 and WR31 (see text for details), compared to V444\,Cyg \citep{lom15}.}
\label{param}
\begin{tabular}{lccc}
  \hline
Parameter & WR21 & WR31 & V444\,Cyg\\
\hline
Sp. Types & WN5+O7V & WN4+O8V & WN5+O6V \\
$R_{\rm WR}$ (R$_{\odot})$ & 4.95 & 2.99 & 2.9 \\
$v_{\infty, \rm WR}$ (\kms) & 1485 & 1713 & 2500 \\
$\dot M_{\rm WR}$ (M$_{\odot}\,$yr$^{-1})$ & $10^{-5}$/$2\times10^{-5}$ &  $2\times10^{-6}$/$2\times10^{-5}$ & $6.8\times10^{-6}$\\
$R_{\rm O}$ (R$_{\odot})$ & 9.37 & 8.52 & 6.85\\
$i (^{\circ})$ & 68.5/49.6/55.0 & 61.7/47.2 & 78.3 \\
$a$ (R$_{\odot})$ & 57.72/70.52/65.56 & 28.85/34.62 & 35.97 \\
        \hline
\end{tabular}
\end{table}

\begin{figure*}
  \begin{center}
    \includegraphics[width=8.5cm]{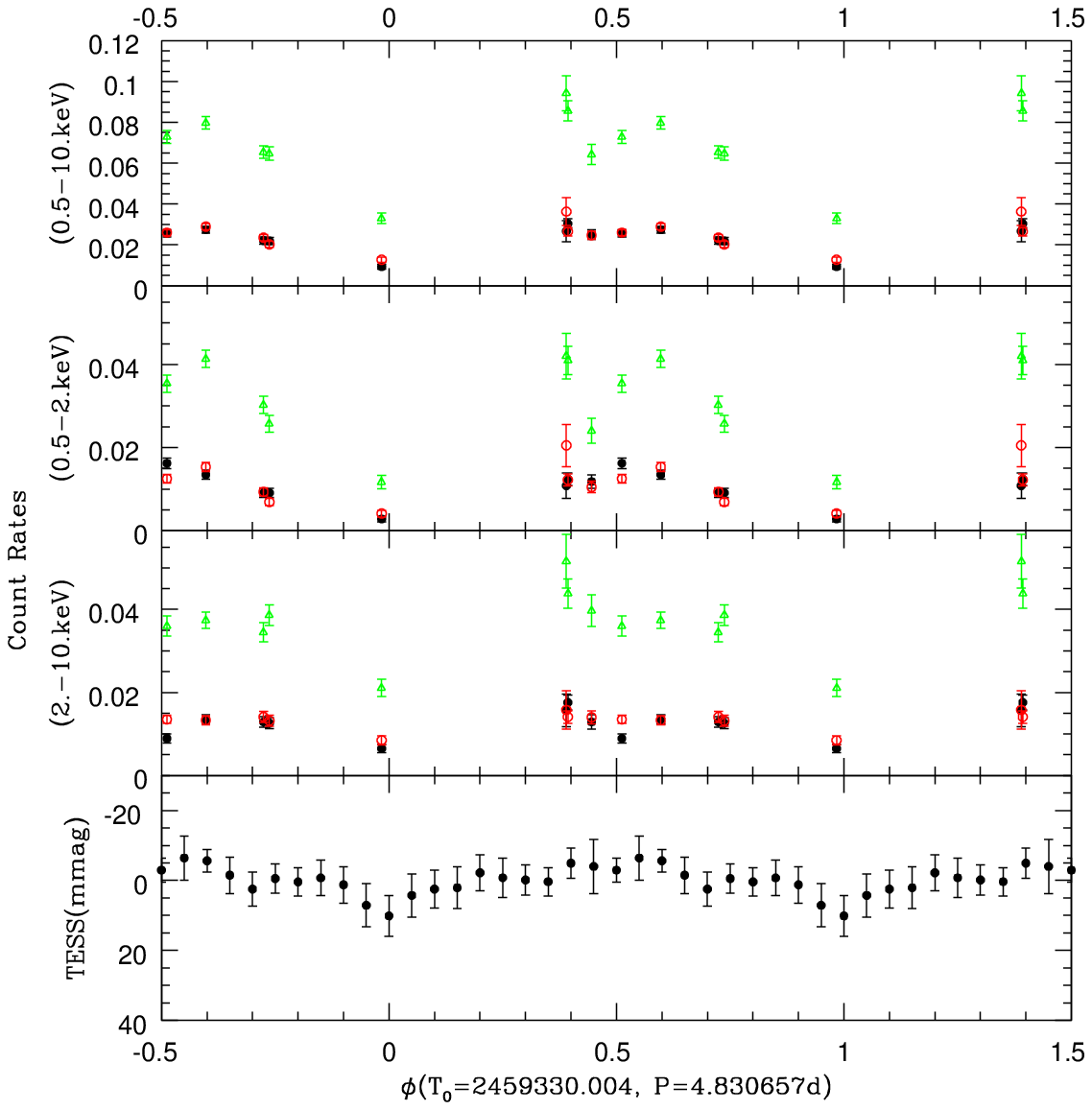}
    \includegraphics[width=8.5cm]{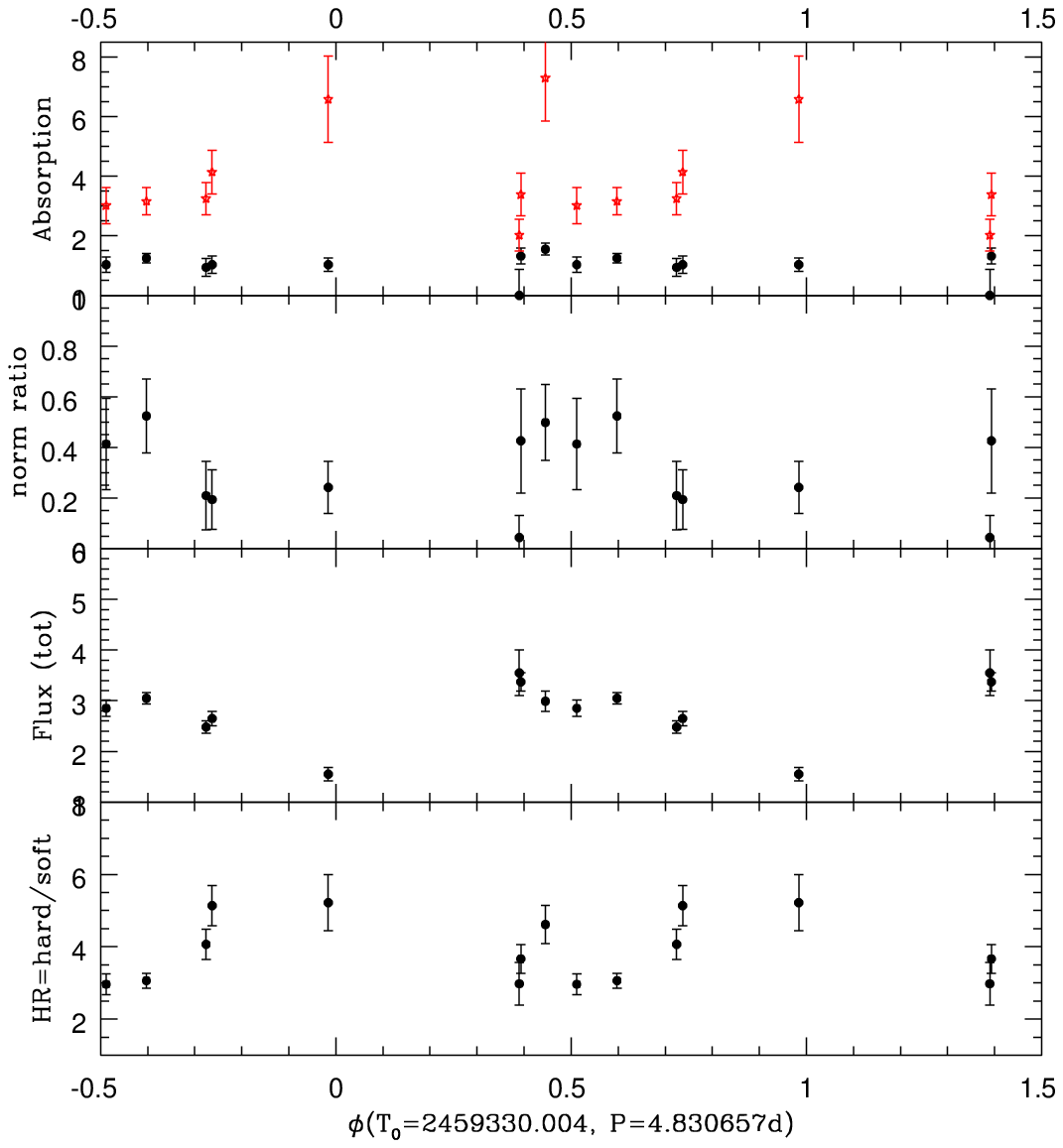}
  \end{center}
  \caption{Left: \xmm\ count rates of WR31 in the three energy bands, compared to {\it TESS} average photometry. EPIC-pn, MOS1, MOS2 data are shown with green triangles, red open squares and black dots, respectively. Right: Results from the spectral fitting using a model $phabs(ISM)\times [ phabs \times apec(0.6) + phabs \times apec(2.0)]$ with solar abundances. From top to bottom are shown the absorptions (filled circles for the coolest component, red stars for the warmest one), the normalization ratio between thermal components $norm(0.6)/norm(2.0)$, the observed flux in total band, and the ratio $HR$ between the hard observed X-ray flux and the soft one. }
\label{wr31cr}
\end{figure*}

The left panel of Fig. \ref{wr21mod} provides a few examples of modelled X-ray light curves in the soft band, normalized by their maximum. Decreasing inclination or mass-loss rates mostly broadens the peak of the light curve. When an offset is considered, the main effect is to shift the maximum away from $\phi=0.5$. Changing the annulus size or the ring brightness leads to asymmetries in the light curve, although the effect of the $w_{\rm pro}$ parameter is usually small.

Since the actual maximum of the observed soft light curve occurs near $\phi\sim 0.6$, an offset of the emitting ring is clearly required for this band. Second, the flux observed when the WR star is in front ($\phi=0$) appears too high in all cases if all soft X-rays are assumed to come from the colliding winds. If instead we assume that only the soft X-rays above the recorded minimum value come from the colliding winds (see end of Sect. 3.1 and green points in Fig. \ref{wr21mod}), a better agreement is found. A contribution from intrinsic shock emission thus appears necessary. Furthermore, the shape of the observed light curve is incompatible with a small annulus as the predicted peak would then be too narrow. However, it agrees quite well with the predictions for the largest annulus. The best models for the soft band (right panel of Fig. \ref{wr21mod}) thus consider a large annulus with some offset and a centering closer to the O star than the WR star. This conclusion is valid for all inclinations and mass-loss rates.

For the hard band, there is no doubt that all observed flux comes from colliding winds (see end of Sect. 3.1). The flux variations with phase are much more limited, with only a factor 1.5 between minimum and maximum (Tables \ref{donx} and \ref{wr21fit}). Such a small variation is not expected in most models. The models with the lower inclination, lower mass-loss rate, and larger annulus are those that produce the closer flux ratios. However, the predicted light curves are then very symmetric, failing to reproduce the lowest fluxes occurring near $\phi=0.5$ (right panel of Fig. \ref{wr21mod}). Reproducing these low fluxes and the maximum near $\phi\sim 0.6$ rather requires the eclipse of a small annulus with a moderate offset and located close to the O star. However, in this case, the fluxes should be strongly absorbed when the WR is in front at $\phi=0$, which is not seen (right panel of Fig. \ref{wr21mod}).

\begin{figure*}
  \begin{center}
    \includegraphics[width=6.5cm,bb=20 516 580 690, clip]{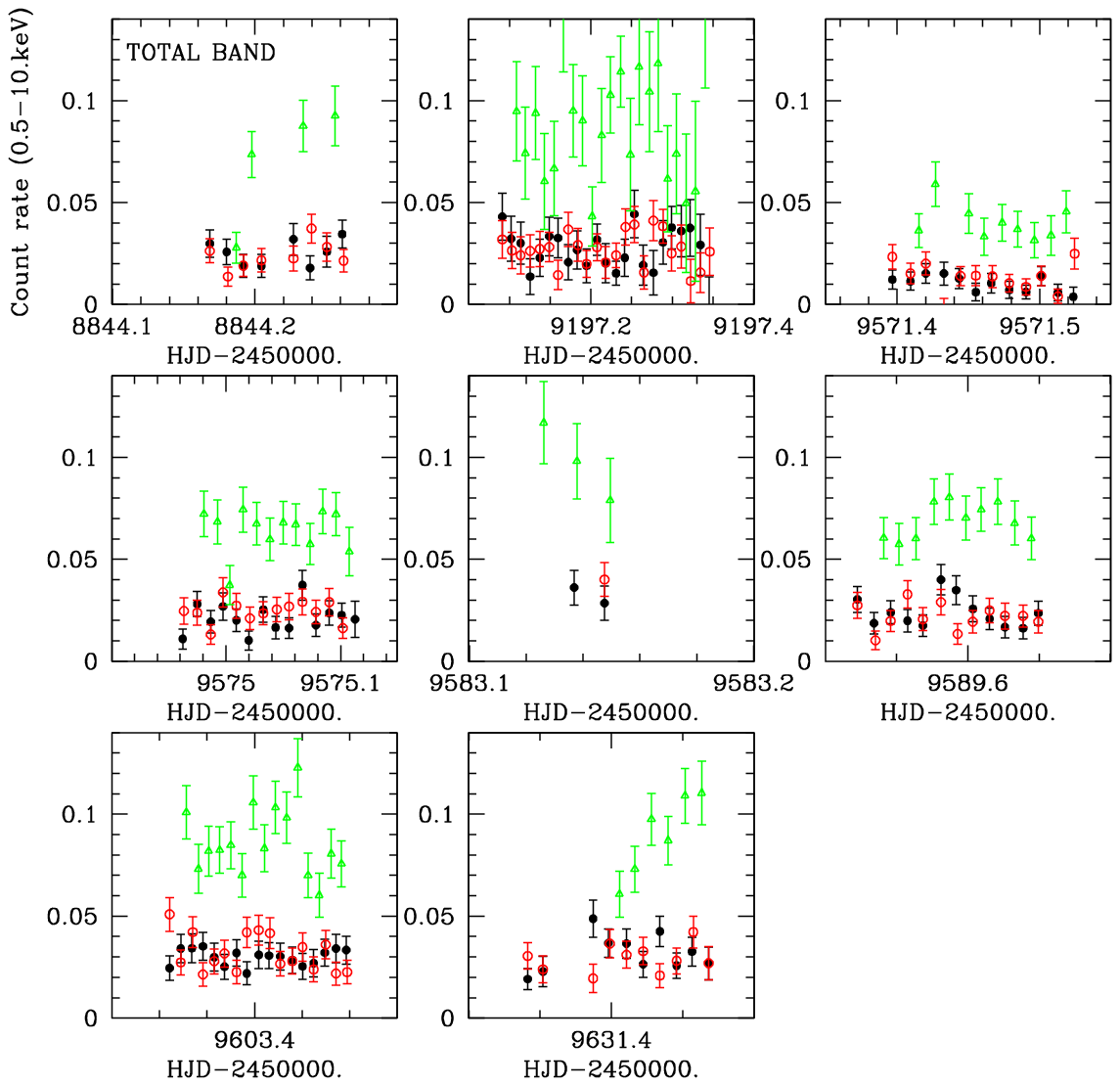}
    \includegraphics[width=6.5cm,bb=20 340 580 516, clip]{WR31_lc_tot.ps}
    \includegraphics[width=4.7cm,bb=20 165 410 340, clip]{WR31_lc_tot.ps}
    \includegraphics[width=6.5cm,bb=20 516 580 690, clip]{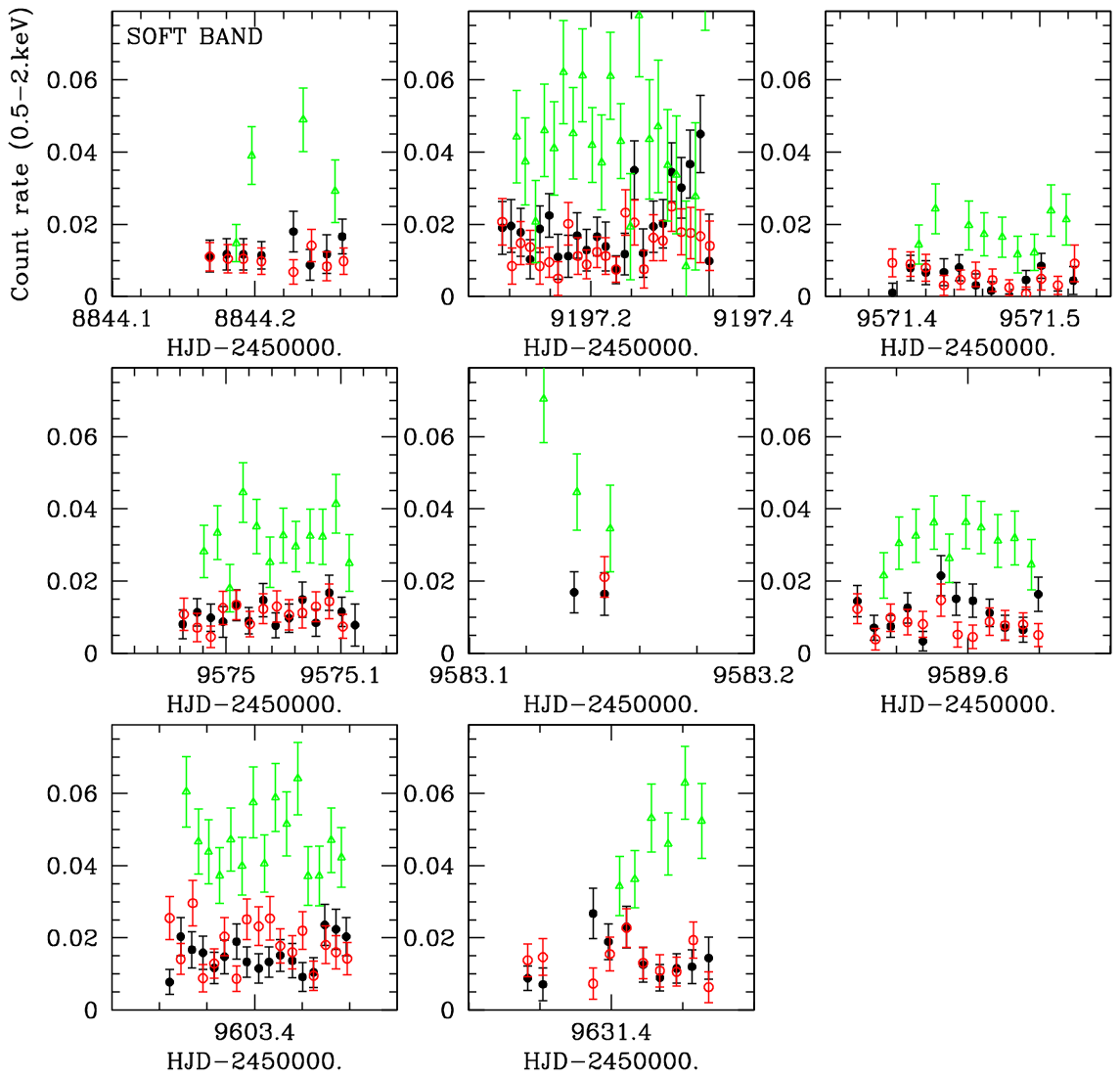}
    \includegraphics[width=6.5cm,bb=20 340 580 516, clip]{WR31_lc_soft.ps}
    \includegraphics[width=4.7cm,bb=20 165 410 340, clip]{WR31_lc_soft.ps}
    \includegraphics[width=6.5cm,bb=20 516 580 690, clip]{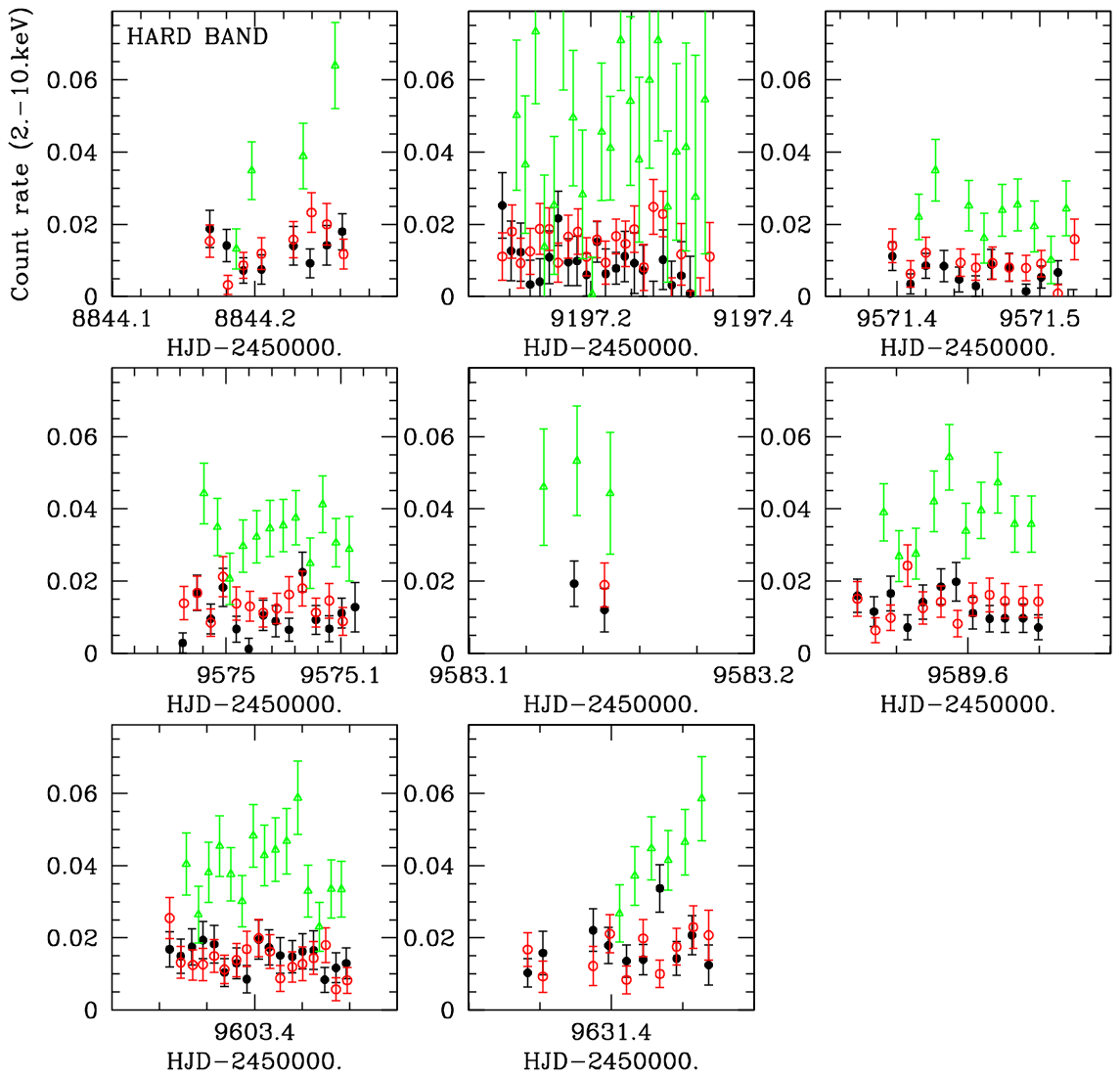}
    \includegraphics[width=6.5cm,bb=20 340 580 516, clip]{WR31_lc_hard.ps}
    \includegraphics[width=4.7cm,bb=20 165 410 340, clip]{WR31_lc_hard.ps}
  \end{center}
  \caption{Same as Fig. \ref{wr21lc} but for WR31.}
\label{wr31lc}
\end{figure*}

\begin{table*}
  \scriptsize
  \caption{Best-fit models to the X-ray spectra of WR31, similar to Table \ref{wr21fit}. \label{wr31fit}}
  \begin{tabular}{lcccccccccccc}
    \hline
\multicolumn{13}{l}{Solar abundances, $phabs(ISM)\times [ phabs \times apec(0.6) + phabs \times apec(2.0)]$, first column fixed to $0.41\times10^{22}$\,cm$^{-2}$}\\
ObsID & $\phi$ & $N_{\rm H}$ & $norm$(0.6\,keV) & $N_{\rm H}$ & $norm$(2.0\,keV) & $\chi^2$/dof & \multicolumn{3}{c}{$F_{\rm X}^{\rm obs}$($10^{-13}$\,erg\,cm$^{-2}$\,s$^{-1}$)} & \multicolumn{3}{c}{$F_{\rm X}^{\rm ISM-cor}$($10^{-13}$\,erg\,cm$^{-2}$\,s$^{-1}$)}  \\
     & & ($10^{22}$\,cm$^{-2}$) & ($10^{-4}$\,cm$^{-5}$) &($10^{22}$\,cm$^{-2}$) & ($10^{-4}$\,cm$^{-5}$) & & tot & soft & hard& tot & soft & hard\\
\hline
0840210401 & 0.45 &1.55$\pm$0.20 &5.03$\pm$1.34 &7.30$\pm$1.45 &10.1$\pm$1.4  &45.06/48   &2.99$\pm$0.20 &0.53$\pm$0.05 &2.46$\pm$0.19 & 3.32$\pm$0.22 & 0.78$\pm$0.07 & 2.53$\pm$0.20 \\
0861710201 & 0.51 &1.03$\pm$0.26 &2.65$\pm$1.12 &3.02$\pm$0.61 &6.41$\pm$0.66 &95.91/82   &2.85$\pm$0.16 &0.72$\pm$0.05 &2.13$\pm$0.16 & 3.31$\pm$0.19 & 1.10$\pm$0.07 & 2.22$\pm$0.19 \\
0880001001 & 0.98 &1.03$\pm$0.22 &1.28$\pm$0.50 &6.58$\pm$1.45 &5.29$\pm$0.91 &35.98/32   &1.55$\pm$0.13 &0.25$\pm$0.03 &1.30$\pm$0.12 & 1.75$\pm$0.15 & 0.40$\pm$0.05 & 1.34$\pm$0.12 \\
0880000901 & 0.72 &0.94$\pm$0.30 &1.31$\pm$0.84 &3.25$\pm$0.54 &6.25$\pm$0.50 &84.39/85   &2.48$\pm$0.12 &0.49$\pm$0.04 &1.99$\pm$0.11 & 2.80$\pm$0.14 & 0.73$\pm$0.06 & 2.07$\pm$0.11 \\
0880000801 & 0.39 &0.00$\pm$0.87 &0.33$\pm$0.66 &2.02$\pm$0.53 &7.57$\pm$1.38 &5.48/12    &3.55$\pm$0.45 &0.89$\pm$0.11 &2.65$\pm$0.41 & 4.28$\pm$0.54 & 1.50$\pm$0.19 & 2.77$\pm$0.43 \\
0880000701 & 0.74 &1.03$\pm$0.29 &1.46$\pm$0.87 &4.14$\pm$0.73 &7.52$\pm$0.73 &93.05/79   &2.65$\pm$0.14 &0.43$\pm$0.04 &2.22$\pm$0.12 & 2.94$\pm$0.16 & 0.64$\pm$0.06 & 2.30$\pm$0.12 \\ 
0880001201 & 0.60 &1.25$\pm$0.15 &3.64$\pm$0.99 &3.16$\pm$0.46 &6.95$\pm$0.45 &126.02/143 &3.05$\pm$0.11 &0.75$\pm$0.04 &2.30$\pm$0.11 & 3.50$\pm$0.13 & 1.10$\pm$0.05 & 2.39$\pm$0.11 \\
0880001301 & 0.39 &1.32$\pm$0.27 &3.51$\pm$1.66 &3.39$\pm$0.72 &8.24$\pm$0.80 &50.60/73   &3.37$\pm$0.18 &0.72$\pm$0.06 &2.65$\pm$0.18 & 3.79$\pm$0.20 & 1.04$\pm$0.09 & 2.75$\pm$0.19 \\
\hline
\multicolumn{13}{l}{WNE abundances, $phabs(ISM)\times [ vphabs \times vapec(0.6) + vphabs \times vapec(2.0)]$, first column fixed to $0.41\times10^{22}$\,cm$^{-2}$}\\
ObsID & $\phi$ & $N_{\rm H}$ & $norm$(0.6\,keV) & $N_{\rm H}$ & $norm$(2.0\,keV) & $\chi^2$/dof & \multicolumn{3}{c}{$F_{\rm X}^{\rm obs}$($10^{-13}$\,erg\,cm$^{-2}$\,s$^{-1}$)} & \multicolumn{3}{c}{$F_{\rm X}^{\rm ISM-cor}$($10^{-13}$\,erg\,cm$^{-2}$\,s$^{-1}$)} \\
     & & ($10^{20}$\,cm$^{-2}$) & ($10^{-5}$\,cm$^{-5}$) &($10^{20}$\,cm$^{-2}$) & ($10^{-5}$\,cm$^{-5}$) & & tot & soft & hard& tot & soft & hard\\
\hline
0840210401 & 0.45 & 3.30$\pm$0.45 &1.53$\pm$0.38 &19.1$\pm$3.84 &3.34$\pm$0.46 & 45.00/48 & 2.97$\pm$0.17 &0.53$\pm$0.04 &2.44$\pm$0.18 & 3.29$\pm$0.19 & 0.77$\pm$0.06 & 2.52$\pm$0.19 \\
0861710201 & 0.51 & 1.98$\pm$0.61 &0.74$\pm$0.35 &7.10$\pm$1.59 &2.08$\pm$0.21 & 97.83/82 & 2.82$\pm$0.16 &0.72$\pm$0.05 &2.10$\pm$0.14 & 3.27$\pm$0.19 & 1.08$\pm$0.07 & 2.19$\pm$0.15 \\
0880001001 & 0.98 & 2.06$\pm$0.47 &0.38$\pm$0.14 &16.6$\pm$3.77 &1.70$\pm$0.29 & 36.92/32 & 1.53$\pm$0.13 &0.25$\pm$0.03 &1.28$\pm$0.12 & 1.72$\pm$0.15 & 0.40$\pm$0.05 & 1.32$\pm$0.12 \\
0880000901 & 0.72 & 1.80$\pm$0.57 &0.38$\pm$0.21 &7.79$\pm$1.27 &2.02$\pm$0.16 & 84.95/85 & 2.45$\pm$0.10 &0.49$\pm$0.04 &1.96$\pm$0.11 & 2.77$\pm$0.11 & 0.73$\pm$0.06 & 2.04$\pm$0.11 \\
0880000801 & 0.39 & 0.03$\pm$1.55 &0.13$\pm$0.58 &4.79$\pm$3.15 &2.46$\pm$0.64 &  5.71/12 & 3.50$\pm$0.53 &0.89$\pm$0.20 &2.60$\pm$0.44 & 4.23$\pm$0.64 & 1.52$\pm$0.34 & 2.71$\pm$0.46 \\
0880000701 & 0.74 & 2.09$\pm$0.60 &0.45$\pm$0.24 &10.3$\pm$1.86 &2.44$\pm$0.23 & 94.88/79 & 2.61$\pm$0.14 &0.43$\pm$0.04 &2.18$\pm$0.13 & 2.90$\pm$0.16 & 0.64$\pm$0.06 & 2.26$\pm$0.13 \\ 
0880001201 & 0.60 & 2.51$\pm$0.32 &1.05$\pm$0.28 &7.67$\pm$1.17 &2.28$\pm$0.14 &126.76/143& 3.03$\pm$0.10 &0.75$\pm$0.04 &2.28$\pm$0.10 & 3.47$\pm$0.11 & 1.09$\pm$0.06 & 2.37$\pm$0.10 \\
0880001301 & 0.39 & 2.76$\pm$0.52 &1.13$\pm$0.45 &8.73$\pm$1.81 &2.74$\pm$0.26 & 50.51/73 & 3.36$\pm$0.15 &0.72$\pm$0.06 &2.64$\pm$0.16 & 3.78$\pm$0.17 & 1.03$\pm$0.08 & 2.74$\pm$0.17 \\
\hline
  \end{tabular}  
\end{table*}

\section{WR31 (WN4+O8V, $P=4.830657$\,d)}

\subsection{X-ray light curves and spectra}
Comparing observations of WR31 (Fig. \ref{wr31cr}), the count rate in the hard band appears quite stable except for the dataset taken at phase 0.0, which lies a factor of two below the other points. In the soft band, the count rate variations are larger (nearly a factor of 4) and more complex. There seems to be a smooth and rather symmetric increase which peaks near $\phi$=0.6 and has a minimum near $\phi$=0.0. This is reminiscent of what is seen for WR\,21, with the peak probably due to absorption changes and the peak offset to Coriolis deflection. On top of that, there is a sharp increase in brightness near $\phi$=0.4. The first observation (ObsID=0880000801) taken at that phase was strongly impacted by background flares hence suffers from a large amount of noise and could be considered somewhat uncertain (despite the care in filtering flares and background choice). However, the second observation (ObsID=0880001301) does not have this problem and appears fully in line with the first, thereby confirming the increase and its phase-locked nature. While narrow flaring events have been detected in $\eta$\,Carinae \citep{mof09}, those flares do not display a repeatable pattern \citep{esp22}. The feature detected in WR31 is therefore of a new type.

Individual light curves of WR31 reveal more variations than for the light curve of WR21 (Fig. \ref{wr31lc}). There is no trace of a short drop which could be associated to an eclipse, but the datasets near $\phi$=0.4--0.45 reveal monotonic trends in EPIC-pn data. The trends are increases for ObsIDs 0840210401 and 0880001301, and a decrease for ObsID 0880000801. The latter two exposures were taken at the same phases ($\phi=0.39$) and it must be underlined that individual points remain compatible, considering the errors, despite different overall trends. Considering the full light curve shape, this can be understood as a narrow peak before the main one, but only a longer exposure covering $\phi=0.35-0.5$ would reveal without ambiguity the exact shape of the variations.

All spectra of WR31 were fitted in a similar way as for WR\,21 (Sect. 3.1). Again, a single temperature component was clearly not sufficient to reproduce all spectra well. As for WR\,21, the very soft and very hard parts of the spectra do not seem to change much and therefore we turned to a model with two-temperature components suffering different absorptions. As the temperatures were not varying much, they were fixed to their average values (0.6 and 2.\,keV) and fits were re-done (Table \ref{wr31fit} and Fig. \ref{wr31cr}). The addition of a third thermal component appeared unnecessary. Using typical WR abundances yields similar trends as with solar abundances (Table \ref{wr31fit}).

While the absorption of the coolest component remains rather stable, the absorption of the hottest component clearly increases (more than doubles) when the WR is in front. As for WR\,21, the strength of the coolest component appears larger at the maximum brightness ($\phi$=0.6). Apart from larger normalizations for both thermal components, the sharp peak observations near $\phi=0.4$ do not correspond to specific changes in either temperature or absorption. 

As for WR\,21, we have calculated the fluxes predicted for embedded wind shocks. This time, the bolometric luminosities $\log[L_{\rm BOL}/L_{\odot}]$ are 4.90 for an O8V secondary star \citep{mar05} and 5.66 for a typical WN4 star \citep{ham19}. For a distance of 8259\,pc \citep{bai21}, X-ray fluxes of 1.3 and $0.01\times 10^{-14}$\,erg\,cm$^{-2}$\,s$^{-1}$ are predicted for the 0.5--2.5\,keV, and 2.5--10.\,keV bands, respectively. The observed X-ray fluxes, after correction by the interstellar absorption, are $6-21$ and $12-22\times 10^{-14}$\,erg\,cm$^{-2}$\,s$^{-1}$ in the 0.5--2.5\,keV, and 2.5-10.0\,keV bands. Again, the observed level of flux in the hard band cannot be explained by the intrinsic emission of the stars. For the soft band, even the minimum observed flux is already four times higher than our (optimistic) estimate, making its origin solely in embedded wind shocks quite unlikely. 

\begin{figure}
  \begin{center}
    \includegraphics[width=8.5cm]{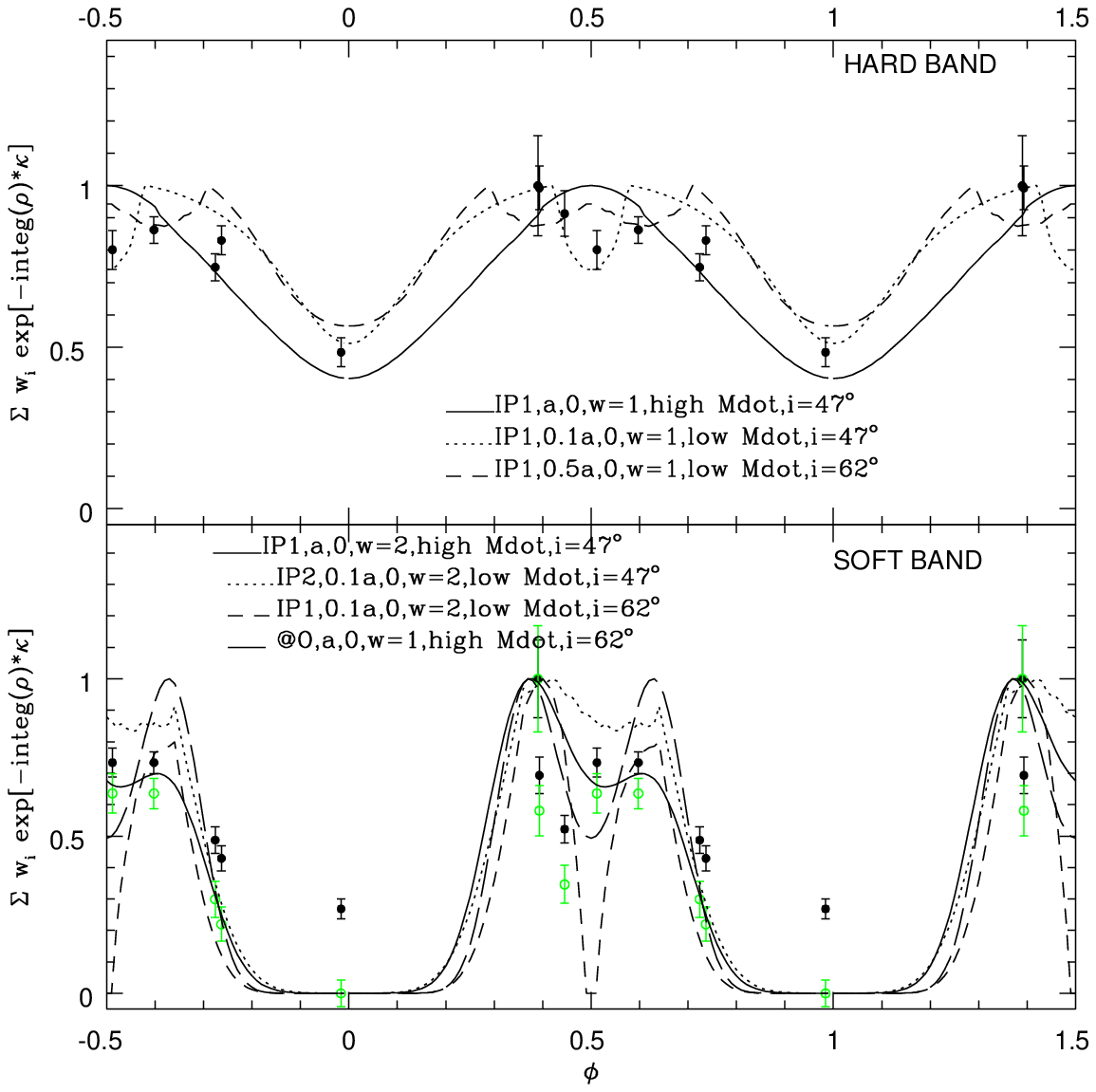}
  \end{center}
  \caption{Best models for WR31, superimposed to the hard (top) and soft (bottom) fluxes.  }
\label{wr31mod}
\end{figure}

\subsection{A simple model}

We ran the same modelling as for WR\,21, but with the orbital and stellar parameters specific to WR\,31 (Table \ref{param}): typical properties of a WN4 star \citep{ham19}, O-star radius of 8.52\,R$_{\odot}$ \citep{mar05}. A lower WR mass-loss rate of $2\times 10^{-6}$\,M$_{\odot}$\,yr$^{-1}$, proposed by \citet{lam96}, was also considered. Regarding inclination, \citet{fah} proposed a value of 61.7$^{\circ}$, resulting in a separation $a=28.85$\,R$_{\odot}$, while our polarimetric data favor a lower value of 47.2$^{\circ}$ (Johnson et al., in prep), leading to a larger separation $a=34.62$\,R$_{\odot}$. 

In the case of WR31, the results are much more dependent on the choice of inclination than for WR21 (Fig. \ref{wr31mod}). This is notably due to eclipses occurring near $\phi=0.5$, which are more pronounced if the inclination is larger. These eclipses can explain the peaks detected at $\phi=0.4$ and 0.6 in the soft band if they are not too narrow. This is confirmed by the poor fit by modelled light curves with offsets. For the low mass-loss rate, the best models are those with a large inclination, a small annulus located slightly closer to the O star than the WR star, and with uniform or moderate leading brightness. For the high mass-loss rate, only models with large annuli can work, with a location of the emission region closer to the O star if the inclination is larger. In both cases, some contribution from the intrinsic X-ray emission of the stars is required to explain the soft X-ray flux seen when the WR is in front ($\phi=0$). 

For the hard band (which is not contaminated by intrinsic emission), the flux appears quasi constant near $\phi=0.4-0.7$ and is slightly smaller at $\phi=0$ (Fig. \ref{wr31mod}). Models combining a high mass-loss rate and a large annulus or a low mass-loss rate with a small annulus provide good results for the lower value of the inclination. Using the larger inclination value, a small annulus would lead to broad eclipses at $\phi=0.5$, which are not seen, while a large annulus would lead to too large variations for the higher mass-loss rate and too small variations for the lower mass-loss rate. A medium-size annulus can work, but only for the low mass-loss rate. 

\section{Discussion and conclusion}

Three short-period massive binaries composed of a WR star and an O star have been monitored in X-rays: WR21, WR31, and V444\,Cyg. All three systems have circular orbits, so that separation changes play no role. The main variations should be linked to absorption, which can be particularly well detected in the soft X-ray band. For all three systems, we indeed observe the largest variations at low energies, about a factor of 3--4. The flux appears maximum when the O star comes into view, and minimum when the WR star and its dense wind cross the line-of-sight. The maximum is not exactly at the conjunction ($\phi=0.5$), though. The flux peaks slightly after conjunction, near $\phi=0.6$. This shifted peak has been interpreted as Coriolis deflection of the collision zone in V444\,Cyg \citep{lom15}, and we can certainly extend this conclusion to WR21. For both stars, the light curve can be well fitted by a large annular emission zone with its center slightly offset in the opposite direction of the O star motion. The situation appears more complicated for WR31, as it presents a second peak near $\phi=0.4$. The shape of that peak is less well defined, since no data were gathered in $\phi=0.-0.4$, impairing interpretation. Note that the soft X-rays are most probably not coming from the collision zone only: a better fit is achieved if one considers the minimum flux to be mostly intrinsic emission from the winds of the stars.

In the hard band, absorption effects should be very limited, but occultations can occur. Indeed, the hard X-rays should come from a smaller region of the collision zone hence they are more prone to eclipse effects. The light curve of V444\,Cyg indeed displayed two clear eclipses of the emission zone, one by the WR star, and one by the O star \citep{lom15}. They correspond to flux decreases by a factor of 4. In contrast, WR21 displays a change by a factor of only 2/3 in flux. Such limited changes are difficult to fit with a very small emission zone, although a variation near $\phi=0.4-0.5$ could be linked to a partial eclipse in such a configuration. For WR31, the recorded variation amounts to a factor of 2, with a minimum when the WR star is in front and a long time of nearly constant flux otherwise: unlike V444\,Cyg, no clear eclipses are seen. This is not surprising since the inclinations of WR21 and WR31 are both much lower than that of V444\,Cyg. Indeed, WR21 and WR31 present only shallow atmospheric eclipses in the optical domain.

For WR21 and WR31, \citet{fah} fitted the spectral variations in the optical domain by a L\"uhrs model of the collision zone. In the case of WR21, the best-fit is achieved with a Coriolis deflection of about 17$^{\circ}$, which corresponds to an offset $\Delta\phi\sim 0.05$. With its $\Delta\phi\sim 0.1$, the observed X-ray lightcurve favors a slightly larger offset. \citet{fah} also deduced a cone half-opening angle of about 40$^{\circ}$ while our models suggest a larger value, as was also found for V444\,Cyg. For WR31, \citet{fah} calculated a deflection by a small amount (8$^{\circ}$) and an even narrower cone (30$^{\circ}$ opening angle). Again, the X-rays favor larger values. In an upcoming paper, we will analyze the spectropolarimetric data, to derive the shape of the collision zone and examine whether it correlates more with the X-ray solution or with the optical emission line solution.

In this paper, we have established the existence of phase-locked X-ray variations in two short-period WN+O binary systems. Our analysis shows that most of the X-rays arise in a colliding-wind interaction. Their variations could be explained by absorption changes along the line-of-sight. The X-ray light curve of such systems yields precious insights into their properties, as done e.g. in V444\,Cyg. However, the detailed analysis of V444\,Cyg could be done because the orbit was extremely well monitored. The current coverage of the orbital cycles of WR21 and WR31 is clearly more limited and it is to be hoped that more X-ray data will be gathered to fill the gaps in the light curves. Only a well sampled light curve will improve our constraints, notably on inclination and mass-loss rates which currently remain entangled especially for WR31, and lead to a better understanding of the colliding wind phenomenon.

\section*{Acknowledgements}
Y.N., G.R., and E.G. acknowledge support from the Fonds National de la Recherche Scientifique (Belgium), the European Space Agency (ESA) and the Belgian Federal Science Policy Office (BELSPO) in the framework of the PRODEX Programme (contracts linked to XMM-Newton). R.A.J. and J.L.H. are grateful for funding from ESA and NASA under AO-20 proposal 88350. J.L.H. also received support from NSF award AST-1816944 and a University of Denver Professional Opportunities for Faculty (PROF) award. The University of Denver resides on the ancestral territories of the Arapaho, Cheyenne, and Ute nations, and its history is inextricably linked with the violent displacement of these indigenous peoples. ADS and CDS were used for preparing this document. 

\section*{Data availability}
The \xmm\ data used in this article are available in the mission's public archives.

\appendix

\section{Additional information}

\begin{figure*}
  \begin{center}
    \includegraphics[width=5.5cm,bb=0 48 739 529, clip]{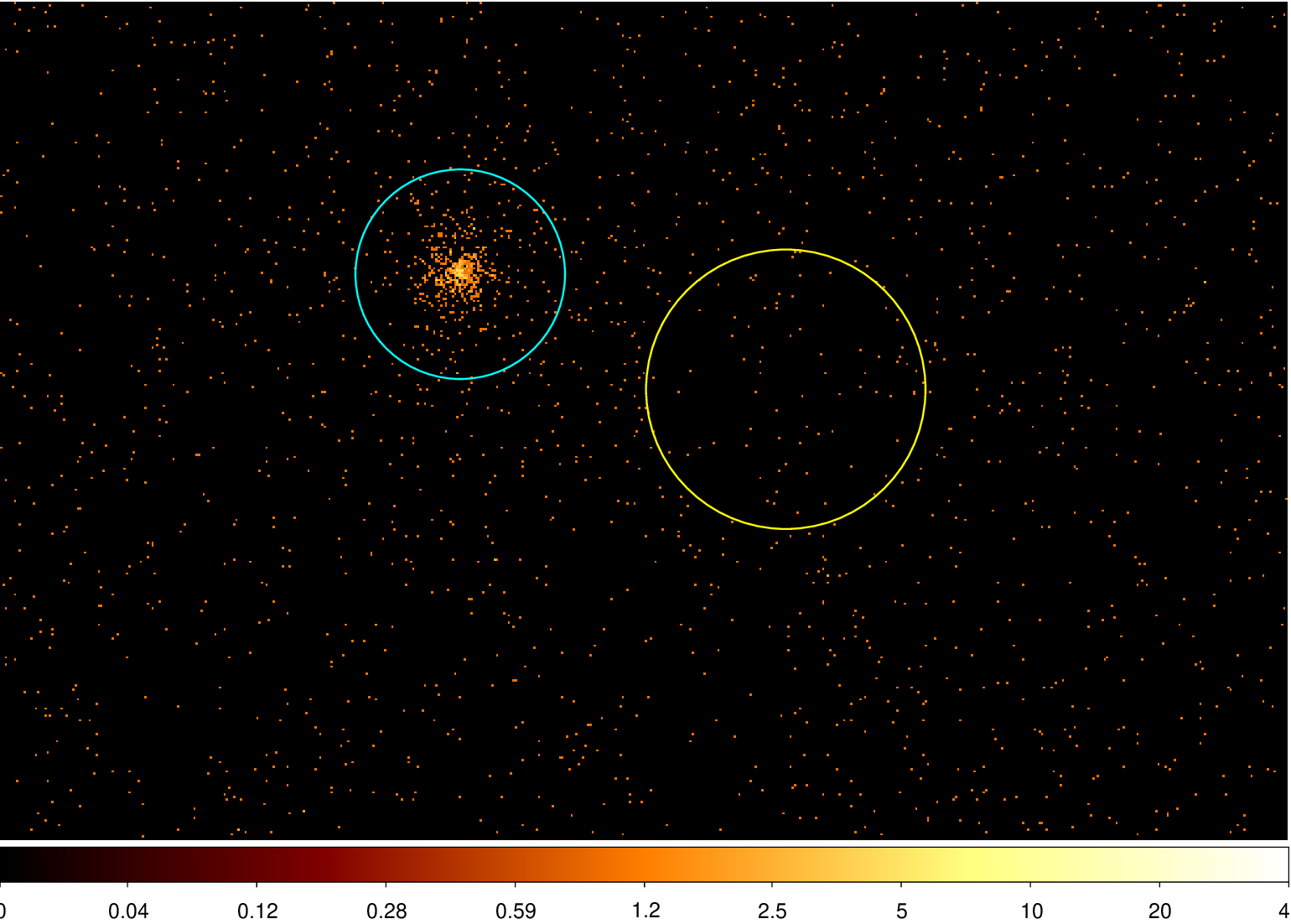}
    \includegraphics[width=5.5cm,bb=0 48 739 529, clip]{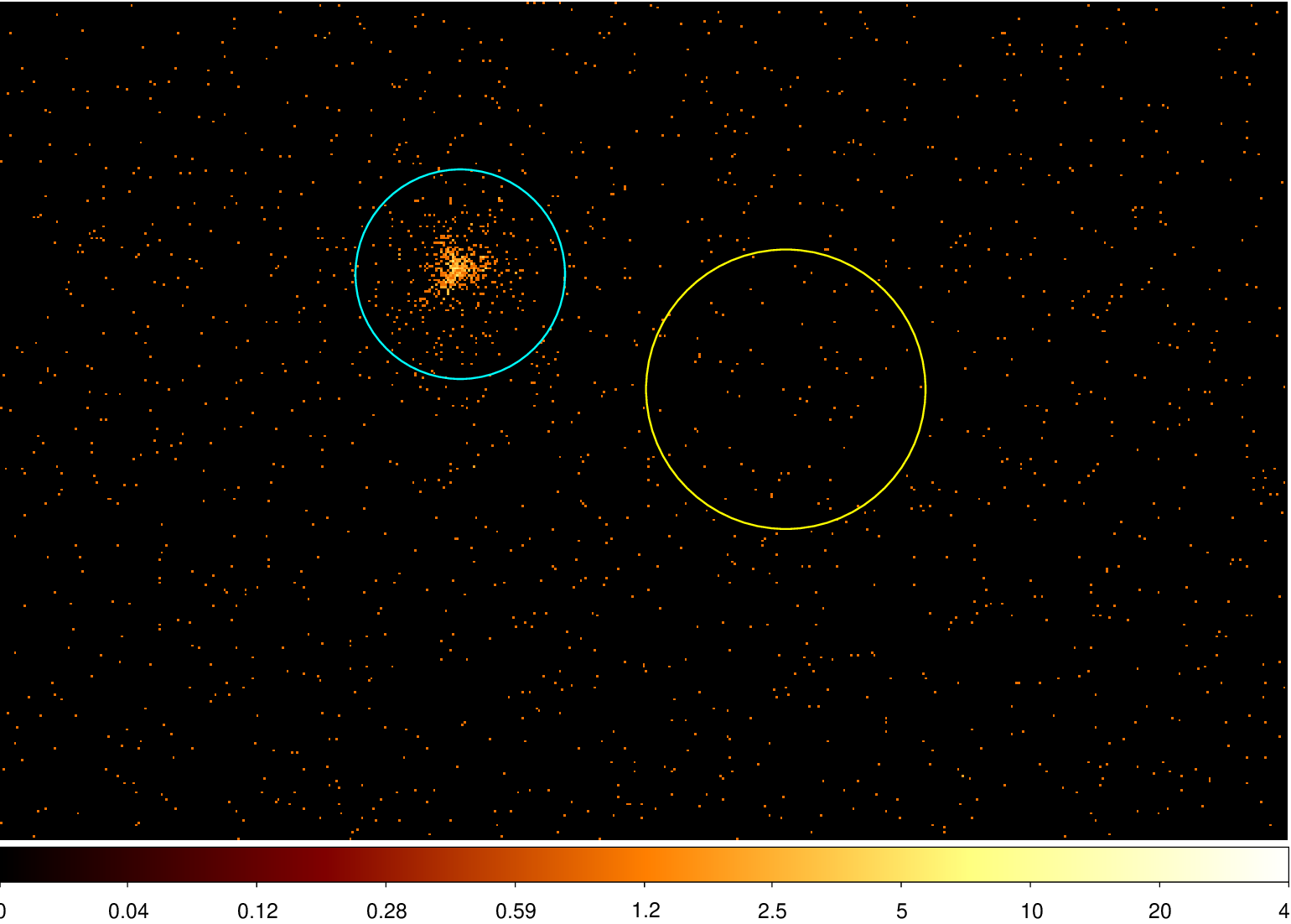}
    \includegraphics[width=5.5cm,bb=0 48 739 529, clip]{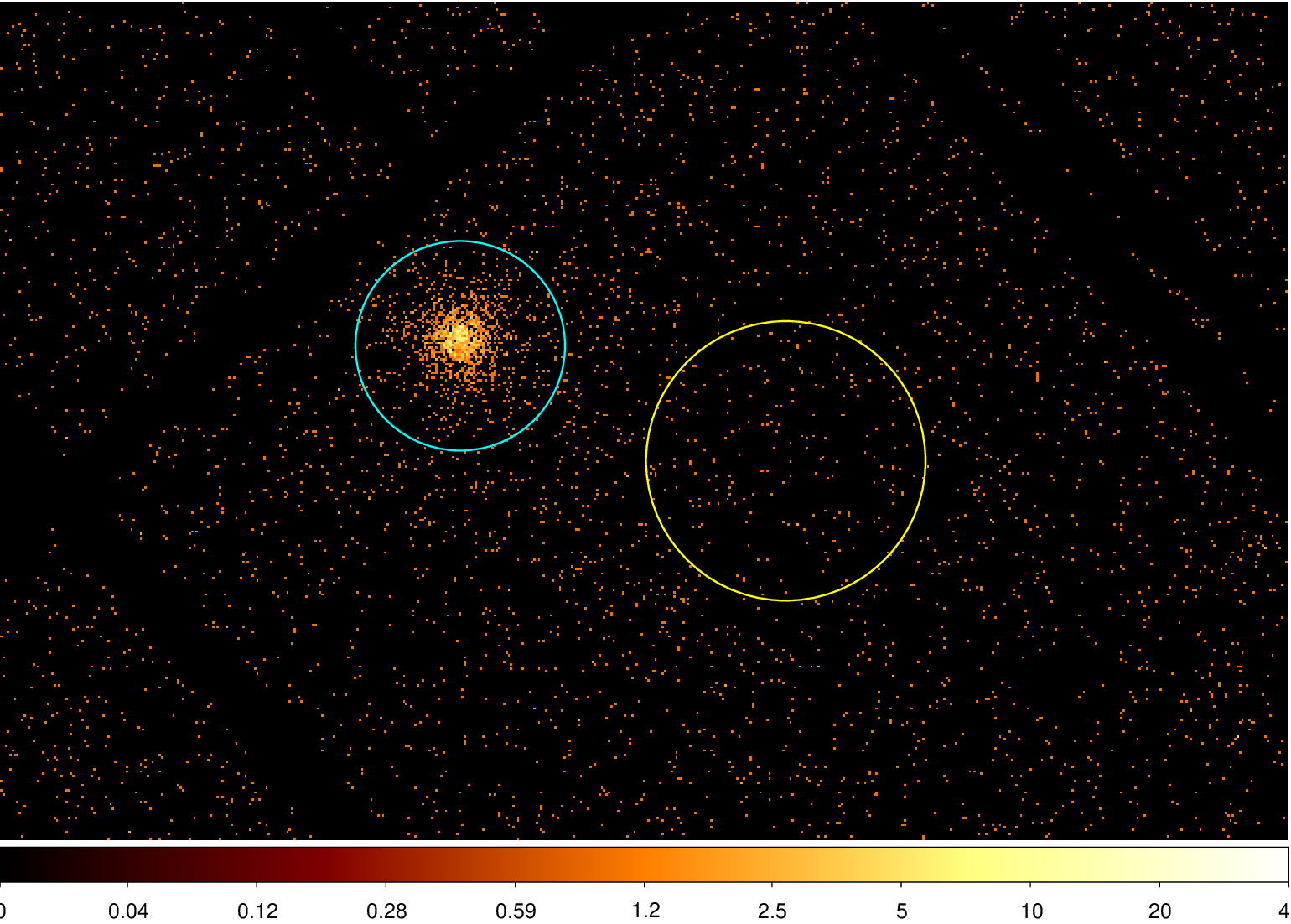}
    \includegraphics[width=5.5cm,bb=0 48 739 529, clip]{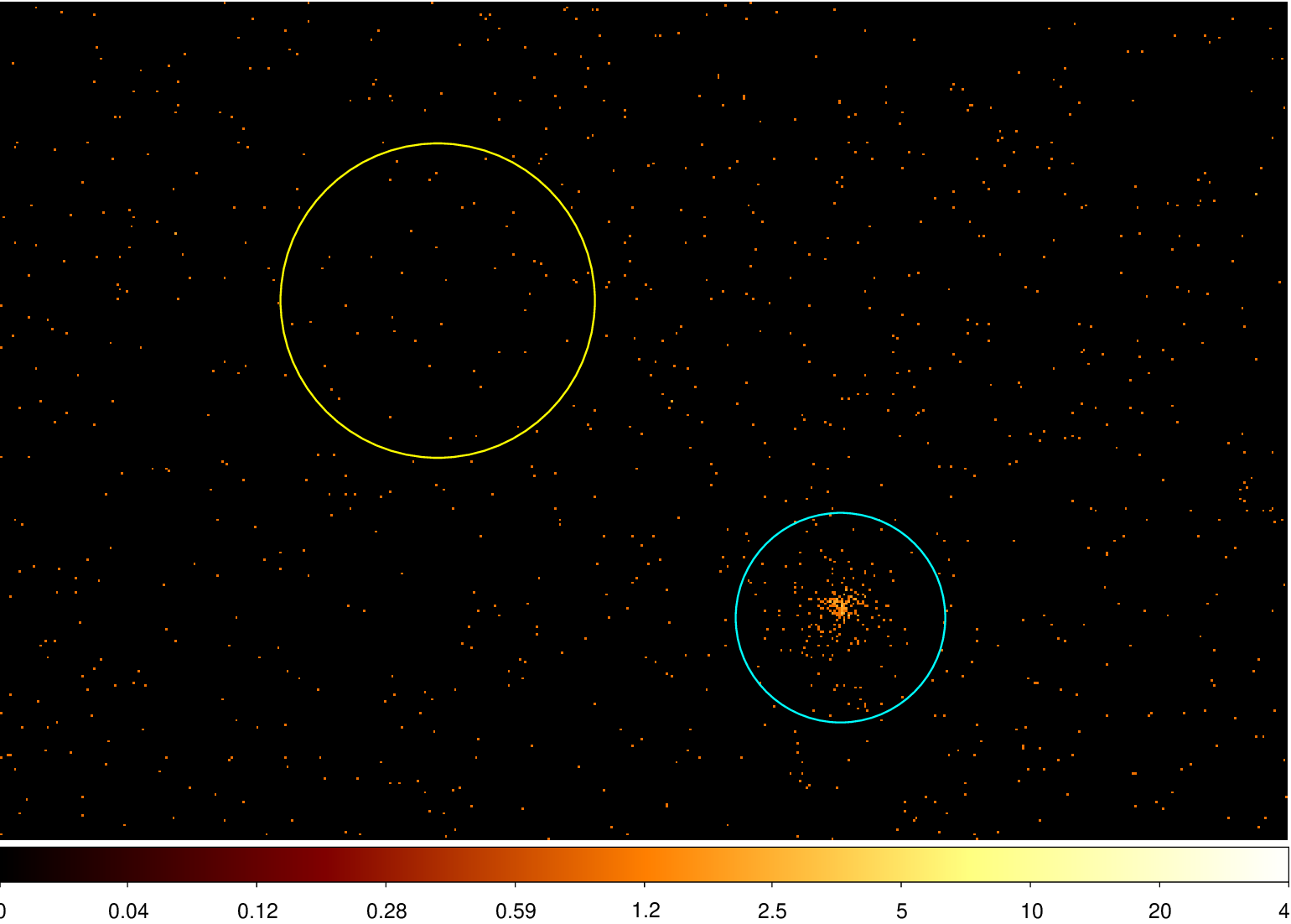}
    \includegraphics[width=5.5cm,bb=0 48 739 529, clip]{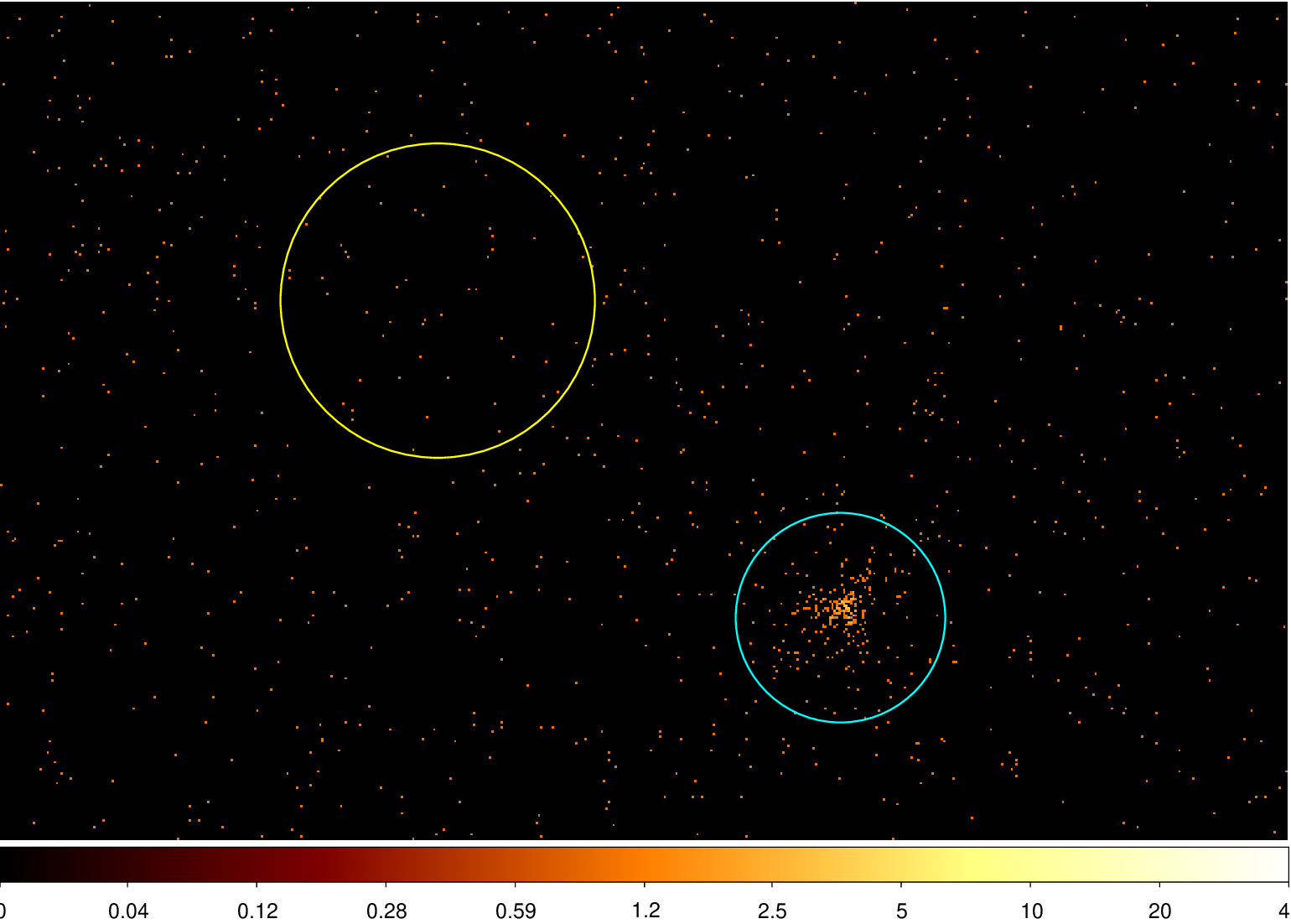}
    \includegraphics[width=5.5cm,bb=0 48 739 529, clip]{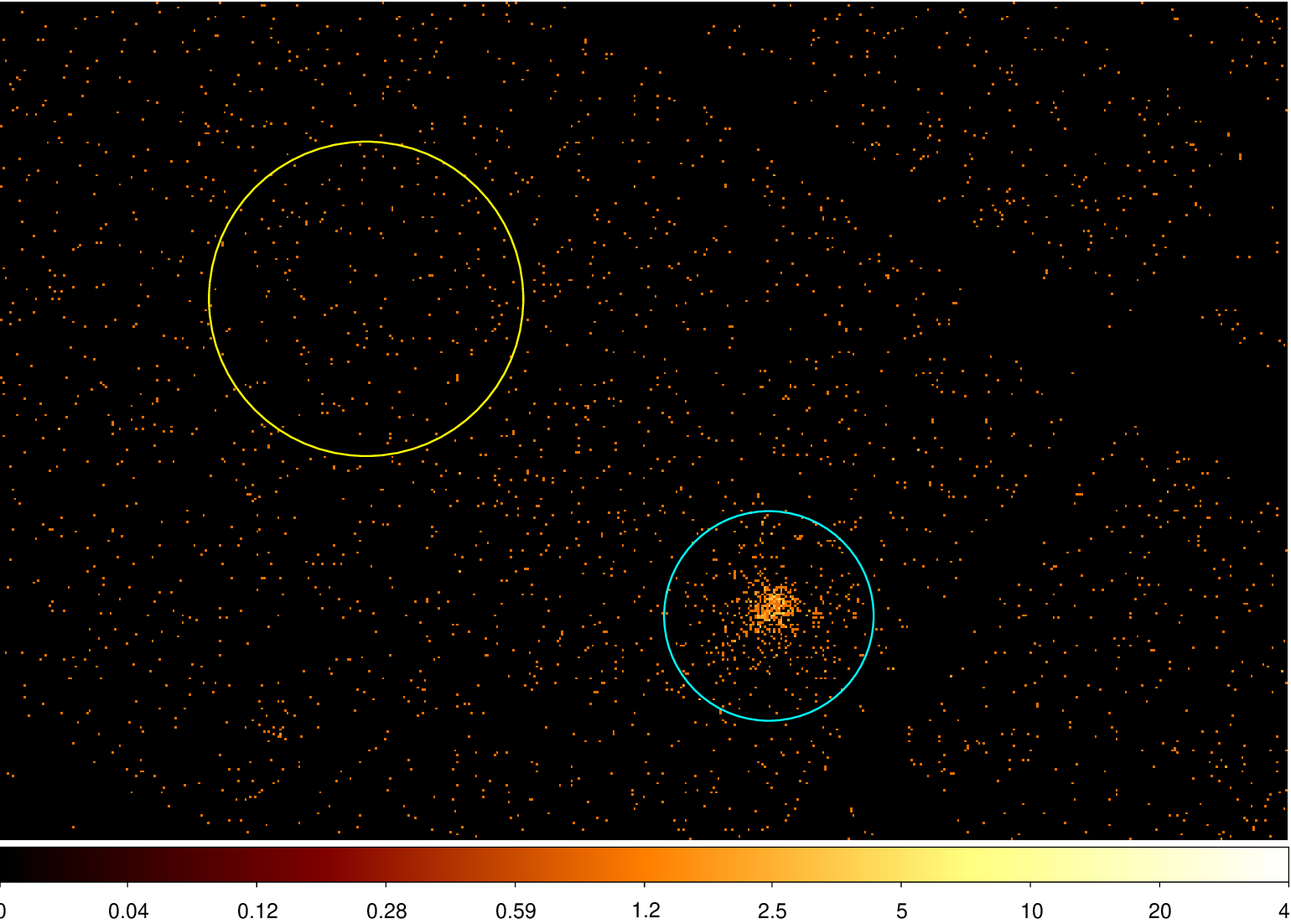}
  \end{center}
  \caption{Source and background regions for WR21 (top, ObsID=0880000401) and WR31 (bottom, ObsID=0880001201), shown for the three cameras (from left to right: EPIC-MOS1, MOS2, and pn) and in the 0.5--10.\,keV band. The regions have radii of 30\arcsec\ for the sources and 40\arcsec\ (WR\,21) or 45\arcsec\ (WR\,31) for the backgrounds. The images, with 0.5\arcsec\ pixels, are shown with a logarithmic scale between 0 and 40 counts.}
\label{wrfig}
\end{figure*}

\begin{figure*}
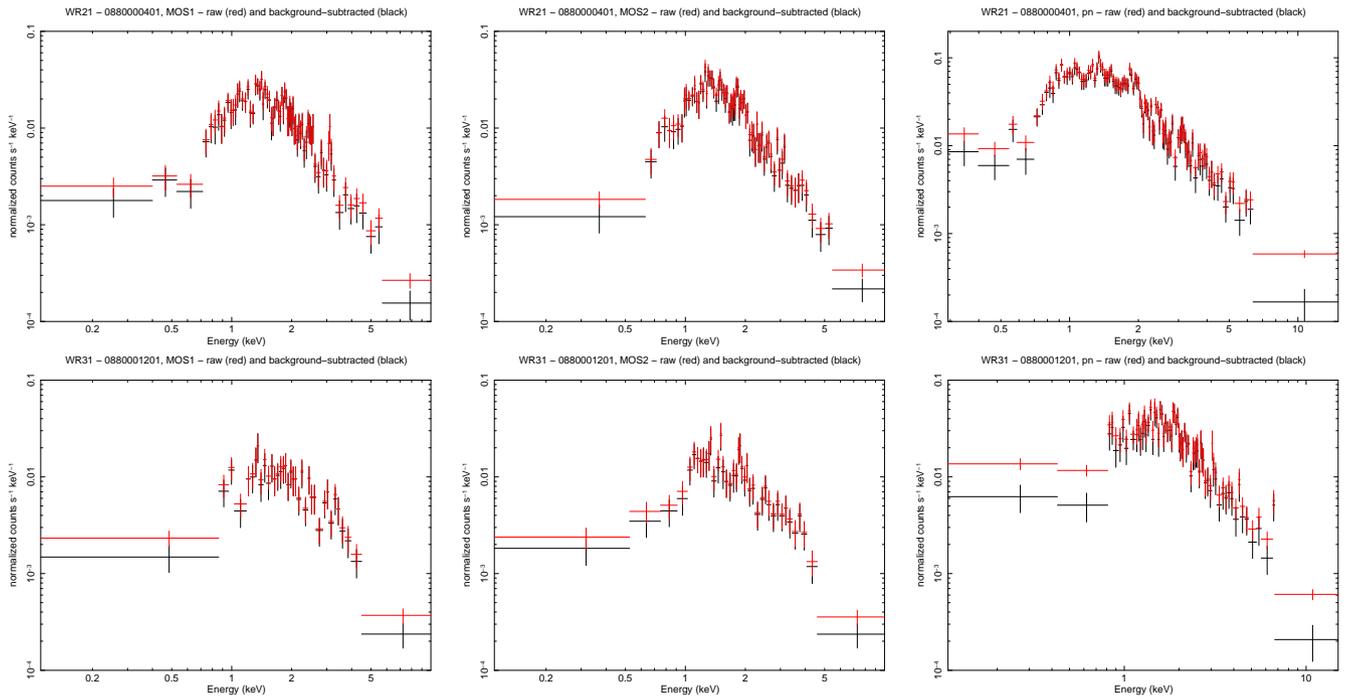

  \begin{center}
    \includegraphics[angle=270,width=5.9cm]{wr21_specm1.ps}
    \includegraphics[angle=270,width=5.9cm]{wr21_specm2.ps}
    \includegraphics[angle=270,width=5.9cm]{wr21_specpn.ps}
    \includegraphics[angle=270,width=5.9cm]{wr31_specm1.ps}
    \includegraphics[angle=270,width=5.9cm]{wr31_specm2.ps}
    \includegraphics[angle=270,width=5.9cm]{wr31_specpn.ps}
  \end{center}
  \caption{Raw (red) and background-corrected (black) spectra of WR21 (top, ObsID=0880000401) and WR31 (bottom, ObsID=0880001201), shown for the three cameras (from left to right: EPIC-MOS1, MOS2, and pn).}
\label{wrspecfig}
\end{figure*}

Figure \ref{wrfig} shows where the source and background were extracted for each target, while Figure \ref{wrspecfig} shows raw and background-corrected spectra for the same exposures. Table \ref{bkgd} also provides the percentage of net counts in the spectral extraction regions. The uniformity of the background and the good agreement between instruments and exposures taken at similar phases confirm the absence of problems due to background subtraction.

  Another fitting trial was made with constant temperature plane-parallel shock plasma ($pshock$) models. For WR\,21, one component was not enough to reach a good fitting in all cases hence two components with individual absorptions were used. The first temperature did not vary significantly hence was fixed to 0.7\,keV. A new fitting then showed that the second temperature remained close to 2.45\,keV and the first absorption to $0.85\times 10^{22}$\,cm$^{-2}$ hence they were also fixed. The final results are provided in Table \ref{fitsh}. The derived fluxes and the absorption trend agree well with those found with the usual $apec$ models (Table \ref{wr21fit}). However, this type of model introduces additional parameters related to the ionization timescale: as can be seen in the Table, they are difficult to constrain and introduce larger errors for other parameters and fluxes. As the WR\,31 data have less counts, the fitting errors would then be even larger in their case so no such trial was done for them.

\begin{table}
  \scriptsize
    \caption{Percentage of net source counts in the raw spectra of each target.}
\label{bkgd}
\begin{tabular}{lccclccc}
  \hline
ObsID & \multicolumn{3}{c}{WR21} & ObsID &\multicolumn{3}{c}{WR31} \\
      & MOS1 & MOS2 & pn & & MOS1 & MOS2 & pn \\
\hline
0780070701 & 89& 91& 81& 0840210401 & 79& 86& 82 \\
0860650501 &   & 80& 64& 0861710201 & 40& 52& 21 \\
0880000301 & 87& 88& 83& 0880001001 & 73& 78& 67 \\
0880000401 & 93& 93& 89& 0880000901 & 84& 86& 81 \\
0880000501 & 89& 90& 83& 0880000801 & 60& 73& 43 \\
0880000601 & 84& 86& 78& 0880000701 & 89& 85& 81 \\ 
           &   &   &   & 0880001201 & 90& 91& 82 \\
           &   &   &   & 0880001301 & 86& 87& 81 \\
        \hline
\end{tabular}
\end{table}

\begin{table*}
  \scriptsize
  \caption{Alternative best-fit models to the X-ray spectra of WR21. \label{fitsh}}
  \begin{tabular}{lcccccccccccc}
    \hline
\multicolumn{13}{l}{Solar abundances, $phabs(ISM)\times [ phabs \times pshock(0.7) + phabs \times pshock(2.45)]$, first two columns fixed to $0.41\times10^{22}$ and $0.85\times 10^{22}$\,cm$^{-2}$}\\
ObsID & $\tau_u^1$ & $norm$(0.7\,keV) & $N_{\rm H}$ & $\tau_u^2$ & $norm$(2.45\,keV) & $\chi^2$/dof & \multicolumn{3}{c}{$F_{\rm X}^{\rm obs}$($10^{-13}$\,erg\,cm$^{-2}$\,s$^{-1}$)}  & \multicolumn{3}{c}{$F_{\rm X}^{\rm ISM-cor}$($10^{-13}$\,erg\,cm$^{-2}$\,s$^{-1}$)}\\
              & ($10^{11}$\,s\,cm$^{-3}$) & ($10^{-4}$\,cm$^{-5}$) & ($10^{22}$\,cm$^{-2}$) & ($10^{11}$\,s\,cm$^{-3}$) & ($10^{-4}$\,cm$^{-5}$) & & tot & soft & hard& tot & soft & hard\\
\hline
0780070701 &17.34$\pm$11.08 &1.82$\pm$0.44 &3.60$\pm$0.67 &0.001$\pm$0.017 &7.92$\pm$0.93 &30.50/41   &2.78$\pm$0.41 &0.66$\pm$0.36 &2.13$\pm$0.27 &3.26$\pm$0.48 &1.05$\pm$0.57 &2.20$\pm$0.28 \\
0860650501 & 3.62$\pm$8.94  &1.00$\pm$0.25 &1.96$\pm$0.36 & 9.84$\pm$7.79  &3.36$\pm$0.23 &74.45/93   &2.26$\pm$0.38 &0.67$\pm$0.30 &1.59$\pm$0.12 &2.67$\pm$0.45 &1.01$\pm$0.45 &1.66$\pm$0.13 \\
0880000301 & 3.96$\pm$4.96  &1.14$\pm$0.21 &2.06$\pm$0.56 & 5.64$\pm$4.29  &3.62$\pm$0.24 &86.46/104  &2.50$\pm$0.28 &0.79$\pm$0.21 &1.71$\pm$0.13 &2.97$\pm$0.33 &1.18$\pm$0.31 &1.78$\pm$0.14 \\
0880000401 & 1.79$\pm$1.94  &2.38$\pm$0.35 &1.61$\pm$0.16 & 4.77$\pm$1.20  &5.11$\pm$0.17 &219.32/239 &4.18$\pm$0.80 &1.63$\pm$0.70 &2.54$\pm$0.15 &5.22$\pm$1.00 &2.56$\pm$1.10 &2.66$\pm$0.16 \\
0880000501 & 0.84$\pm$1.06  &1.15$\pm$0.16 &1.62$\pm$0.26 & 7.22$\pm$3.61  &3.64$\pm$0.18 &128.37/142 &2.68$\pm$0.36 &0.90$\pm$0.25 &1.78$\pm$0.13 &3.27$\pm$0.44 &1.41$\pm$0.39 &1.86$\pm$0.14 \\ 
0880000601 & 500.$\pm$480.  &1.32$\pm$0.12 &5.35$\pm$0.67 &40.13$\pm$29.85 &5.80$\pm$0.47 &133.82/114 &2.48$\pm$1.10 &0.40$\pm$0.33 &2.08$\pm$0.85 &2.81$\pm$1.25 &0.66$\pm$0.54 &2.15$\pm$0.88 \\
\hline
  \end{tabular}
  
{\scriptsize Total, soft and hard energy bands being defined as 0.5--10.0\,keV, 0.5--2.0\,keV, and 2.0--10.0\,keV, respectively. Errors are 1$\sigma$ uncertainties; they correspond to the larger value if the error bar is asymmetric. }
\end{table*}

\bsp	% typesetting comment
\label{lastpage}

\begin{thebibliography}{99}
\bibitem[Asplund et al.(2009)]{asp09} Asplund, M., Grevesse, N., Sauval, A.J., \& Scott, P.\ 2009, \araa, 47, 481   
\bibitem[Bailer-Jones et al.(2021)]{bai21} Bailer-Jones, C.~A.~L., Rybizki, J., Fouesneau, M., et al.\ 2021, \aj, 161, 147. doi:10.3847/1538-3881/abd806
\bibitem[Bestenlehner et al.(2022)]{bes22} Bestenlehner, J.~M., Crowther, P.~A., Broos, P.~S., et al.\ 2022, \mnras, 510, 6133. doi:10.1093/mnras/stab3521
\bibitem[Buckley et al.(2006)]{buc06} Buckley, D.~A.~H., Swart, G.~P., \& Meiring, J.~G.\ 2006, \procspie, 6267, 62670Z. doi:10.1117/12.673750
\bibitem[Cherepashchuk(1976)]{che76} Cherepashchuk, A.~M.\ 1976, Soviet Astronomy Letters, 2, 138
\bibitem[Espinoza-Galeas et al.(2022)]{esp22} Espinoza-Galeas, D., Corcoran, M.~F., Hamaguchi, K., et al.\ 2022, \apj, 933, 136. doi:10.3847/1538-4357/ac69ce
\bibitem[Fahed \& Moffat(2012)]{fah} Fahed, R. \& Moffat, A.~F.~J.\ 2012, \mnras, 424, 1601. doi:10.1111/j.1365-2966.2012.20494.x
\bibitem[Fullard et al.(2022)]{ful22} Fullard, A.~G., O'Brien, J.~T., Kerzendorf, W.~E., et al.\ 2022, \apj, 930, 89. doi:10.3847/1538-4357/ac589e
\bibitem[Gayley et al.(1997)]{gay97} Gayley, K.~G., Owocki, S.~P., \& Cranmer, S.~R.\ 1997, \apj, 475, 786. doi:10.1086/303573
\bibitem[Gosset et al.(2001)]{hmm} Gosset, E., Royer, P., Rauw, G., et al.\ 2001, \mnras, 327, 435. doi:10.1046/j.1365-8711.2001.04755.x
\bibitem[Gosset (2007)]{gos07} Gosset, E.\ 2007, Habilitation Thesis, University of Li\`ege, Belgium
\bibitem[Gosset et al.(2009)]{gos09} Gosset, E., Naz{\'e}, Y., Sana, H., et al.\ 2009, \aap, 508, 805. doi:10.1051/0004-6361/20077981
\bibitem[Gosset \& Naz{\'e}(2016)]{gos16} Gosset, E. \& Naz{\'e}, Y.\ 2016, \aap, 590, A113. doi:10.1051/0004-6361/201527051
\bibitem[Graham et al.(2013)]{gra13} Graham, M.~J., Drake, A.~J., Djorgovski, S.~G., et al.\ 2013, \mnras, 434, 3423. doi:10.1093/mnras/stt1264
\bibitem[Hamann et al.(2019)]{ham19} Hamann, W.-R., Gr{\"a}fener, G., Liermann, A., et al.\ 2019, \aap, 625, A57. doi:10.1051/0004-6361/201834850
\bibitem[Johnson et al.(2019)]{jon19} Johnson, R.~A., Fullard, A.~G., Lomax, J.~R., et al.\ 2019, Research Notes of the American Astronomical Society, 3, 146. doi:10.3847/2515-5172/ab4a12
\bibitem[Kobulnicky et al.(2003)]{kob03} Kobulnicky, H.~A., Nordsieck, K.~H., Burgh, E.~B., et al.\ 2003, \procspie, 4841, 1634. doi:10.1117/12.460315
\bibitem[Lamontagne et al.(1996)]{lam96} Lamontagne, R., Moffat, A.~F.~J., Drissen, L., et al.\ 1996, \aj, 112, 2227. doi:10.1086/118175
\bibitem[Lomax et al.(2015)]{lom15} Lomax, J.~R., Naz{\'e}, Y., Hoffman, J.~L., et al.\ 2015, \aap, 573, A43. doi:10.1051/0004-6361/201424468
\bibitem[Martins et al.(2005)]{mar05} Martins, F., Schaerer, D., \& Hillier, D.~J.\ 2005, \aap, 436, 1049. doi:10.1051/0004-6361:20042386
\bibitem[Moffat \& Corcoran(2009)]{mof09} Moffat, A.~F.~J. \& Corcoran, M.~F.\ 2009, \apj, 707, 693. doi:10.1088/0004-637X/707/1/693
\bibitem[Naz{\'e} et al.(2011)]{naz11} Naz{\'e}, Y., Broos, P.~S., Oskinova, L., et al.\ 2011, \apjs, 194, 7. doi:10.1088/0067-0049/194/1/7
\bibitem[Naz{\'e} et al.(2018)]{naz18} Naz{\'e}, Y., Koenigsberger, G., Pittard, J.~M., et al.\ 2018, \apj, 853, 164. doi:10.3847/1538-4357/aaa29c
\bibitem[Naz{\'e} et al.(2021)]{nazwr} Naz{\'e}, Y., Gosset, E., \& Marechal, Q.\ 2021, \mnras, 501, 4214. doi:10.1093/mnras/staa3801
\bibitem[Niemela et al.(1995)]{nie95} Niemela, V.~S., Cabanne, M.~L., \& Bassino, L.~P.\ 1995, \rmxaa, 31, 45
\bibitem[Panagiotou \& Walter(2018)]{pan18} Panagiotou, C. \& Walter, R.\ 2018, \aap, 610, A37. doi:10.1051/0004-6361/201731841
\bibitem[Pandey et al.(2014)]{pan14} Pandey, J.~C., Pandey, S.~B., \& Karmakar, S.\ 2014, \apj, 788, 84. doi:10.1088/0004-637X/788/1/84
\bibitem[Pollock(1987)]{pol87} Pollock, A.~M.~T.\ 1987, \apj, 320, 283. doi:10.1086/165539
\bibitem[Pollock et al.(2018)]{pol18} Pollock, A.~M.~T., Crowther, P.~A., Tehrani, K., et al.\ 2018, \mnras, 474, 3228. doi:10.1093/mnras/stx2879
\bibitem[Rauw et al.(2000)]{rau00} Rauw, G., Stevens, I.~R., Pittard, J.~M., et al.\ 2000, \mnras, 316, 129. doi:10.1046/j.1365-8711.2000.03491.x
\bibitem[Schwarzenberg-Czerny(1989)]{aov} Schwarzenberg-Czerny, A.\ 1989, \mnras, 241, 153. doi:10.1093/mnras/241.2.153
\bibitem[Stevens et al.(1992)]{ste92} Stevens, I.~R., Blondin, J.~M., \& Pollock, A.~M.~T.\ 1992, \apj, 386, 265. doi:10.1086/171013
\bibitem[Willis et al.(1995)]{wil95} Willis, A.~J., Schild, H., \& Stevens, I.~R.\ 1995, \aap, 298, 549
\end{thebibliography}
\end{document}